\documentclass[%
reprint,
superscriptaddress,
 amsmath,amssymb,
 aps,
]{revtex4-1}

\usepackage{graphicx}
\usepackage{dcolumn}
\usepackage{bm}
\usepackage[colorlinks=true,allcolors=blue]{hyperref}

\usepackage{pgf,tikz}
\usepackage{multirow,booktabs}
\usepackage{textcomp} 
\usetikzlibrary{matrix, arrows, automata}
\usepackage{blkarray} 
\usepackage{esint} 
\usepackage{blindtext}
\usepackage{color}
\usepackage[normalem]{ulem} 

\usepackage{chngcntr} 

\newcommand{\units}[1]{\,\mathrm{#1}}
\newcommand{\chg}[2][red]{{\color{#1}#2}}
   \newcommand{\chgchg}[2][red]{{\color{#1}#2}}
   \newcommand{\chgGreen}[2][red]{{\color{#1}#2}}
\newcommand{\chgrev}[2][red]{{\color{#1}#2}}

\newcommand\rem[1]{}

\renewcommand{\chg}[2][black]{#2}
\renewcommand{\chgchg}[2][black]{#2}
\renewcommand{\chgrev}[2][black]{#2}
\renewcommand{\chgGreen}[2][black]{#2}

\usepackage[pagewise]{lineno}
\begin{document}


\title{\Large{Physical and geometric determinants of transport in\\ feto-placental microvascular networks}}
\thanks{A.E., P.P., O.E.J., I.L.C. designed research, A.E., P.P., R.P.M., O.E.J., I.L.C. performed research and A.E., P.P., O.E.J., I.L.C.  wrote the manuscript.}

\author{Alexander Erlich}
\thanks{A.E. and P.P. contributed equally to this work.}
\affiliation{School of Mathematics, University of Manchester, Oxford Road, Manchester M13 9PL, UK}
\author{Philip Pearce}
\thanks{A.E. and P.P. contributed equally to this work.}
\affiliation{Department of Mathematics, Massachusetts Institute of Technology, 77 Massachusetts Avenue, Cambridge, Massachusetts 02139-4307, USA}
\author{Romina Plitman Mayo}
\affiliation{Centre for Trophoblast Research, Department of Physiology, Development and Neuroscience, University of Cambridge, Cambridge  CB2 3EG UK}
\affiliation{Homerton College, University of Cambridge, Cambridge, CB2 8PH}
\author{Oliver E. Jensen}
\affiliation{School of Mathematics, University of Manchester, Oxford Road, Manchester M13 9PL, UK}
\author{Igor L. Chernyavsky}
\email[To whom correspondence should be addressed. E-mail: ]{igor.chernyavsky@manchester.ac.uk}
\affiliation{School of Mathematics, University of Manchester, Oxford Road, Manchester M13 9PL, UK}
\affiliation{Maternal and Fetal Health Research Centre, Division of Developmental Biology and Medicine, School of Medical Sciences, University of Manchester, Manchester Academic Health Science Centre, Manchester M13 9PL, UK}

\date{\today}

\begin{abstract}
Across mammalian species, solute exchange takes place in complex microvascular networks. In the human placenta, the primary exchange units are terminal villi that contain disordered networks of fetal capillaries and are surrounded externally by maternal blood. Here we show how the irregular internal structure of a terminal villus determines its exchange capacity for a wide range of solutes. Distilling geometric features into three scalar parameters, obtained from image analysis and computational fluid dynamics, we \chg{capture archetypal features} of the the structure-function relationship of terminal villi using a simple algebraic approximation, revealing transitions between flow- and diffusion-limited transport at vessel and network levels. Our theory accommodates countercurrent effects, incorporates nonlinear blood rheology and offers an efficient method for testing network robustness. Our results show how physical estimates of solute transport, based on carefully defined geometrical statistics, provide a viable method for linking placental structure and function, and offer a framework for assessing transport in other microvascular systems. 
\end{abstract}

\pacs{Valid PACS appear here}
\maketitle


The human placenta performs diverse functions later taken on by several different organs \cite{burton2015}. In particular, it mediates the exchange of vital solutes, including respiratory gases and nutrients, between the mother and the developing fetus. The complex heterogeneous structure of the placenta is adapted to perform these various functions. However, despite its availability for \emph{ex vivo} perfusion experiments just after birth, and the importance of placental dysfunction in conditions such as fetal growth restriction, the link between placental structure and function in health and disease remains poorly understood \cite{serov2015role, jensenchernyavsky2018}. Multiscale models have proved successful in investigating aspects of the structure-function relationship in the microcirculation \cite{Secomb_etal13,FrySecomb13}, lymph nodes \cite{Moore15} and organs including the brain \cite{Blinder13,HadjistassouVentikos15,GouldLinninger_etal16,peyrounette2018multiscale}, the kidney \cite{MoralesKalaidzidis15} and the liver \cite{Siggers_etal10,MoralesKalaidzidis15}.  However general methods for incorporating experimental data on complex, heterogeneous capillary networks into such models remain under-developed.

\begin{figure*} 
\centering
\includegraphics[width=1\textwidth]{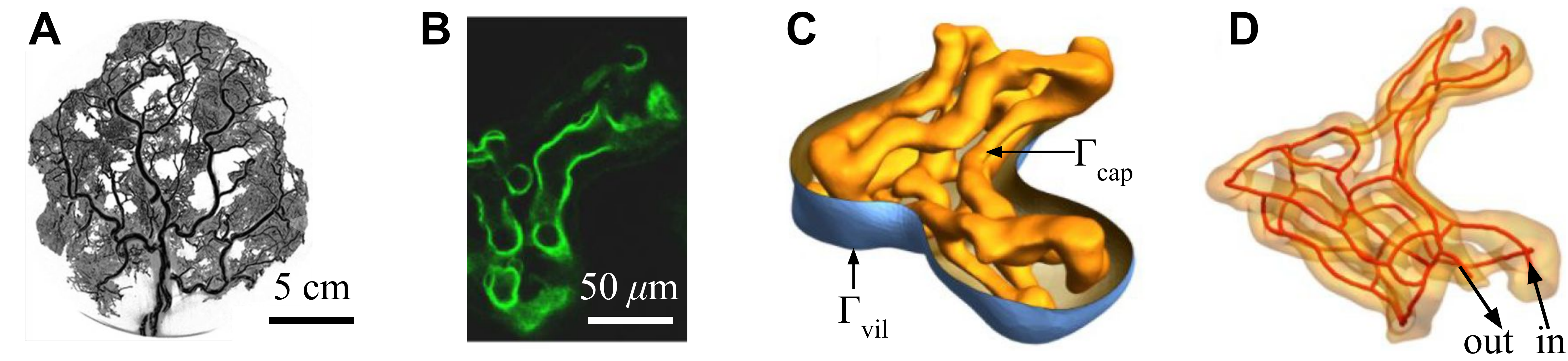}
\caption{The multiscale structure of the feto-placental vasculature.  \textbf{A} Feto-placental arterial vessels (imaged using micro-{X-ray tomography}; reproduced {with permission via CC-BY} from \cite{JunaidJohnstone17})
deliver blood from the umbilicus through numerous bifurcating vessels to peripheral capillary networks (\hbox{e.g.} \textbf{B}, imaged using confocal microscopy).  The feto-placental vasculature is confined within villous trees that are coated with syncytiotrophoblast and are bathed in maternal blood; capillary networks sit within terminal villi, the peripheral branches of the trees.  \textbf{C} A segmented confocal image of a terminal villus reveals the surface $\Gamma_{\text{cap}}$ of fetal capillaries (yellow) and the surrounding syncytiotrophoblast (blue, $\Gamma_{\text{vil}}$) that interfaces with maternal blood.  Image processing yields capillary centerlines (\textbf{D}, red), which have total length $L_\text{c}$. The assumed inlet and outlet vessels are indicated. Fetal blood occupies the volume $\Omega_b$ confined by $\Gamma_{\text{cap}}$; villous tissue occupies the space $\Omega_t$ between $\Gamma_{\text{cap}}$ and $\Gamma_{\text{vil}}$.} 
\label{fig:imaging}
\end{figure*}

Recent advances in three-dimensional (3D) imaging have revealed aspects of placental structure in intricate detail \cite{
jirkovska2008three,mayo2016three,perazzolo2017modelling,JunaidJohnstone17} (Fig.~\ref{fig:imaging}). Fetal blood flows from the umbilical cord through a complex network of vessels that are confined within multiple villous trees; the trees sit in chambers that are perfused with maternal blood.   Much of the solute exchange between maternal and fetal blood takes place across the thin-walled peripheral branches of the trees (terminal villi), which contain the smallest feto-placental capillaries. Quantitative measurements have demonstrated structural differences between healthy and pathological placentas \chgrev{(such as changes in villous capillary network density) \cite{Mayhew04}}, but physical explanations for the observed symptoms of diseases such as pre-eclampsia and diabetes have so far been confined mainly to analyses of diffusive conductances from two-dimensional histological data \cite{Mayhew04, Mayhew_etal07, Rainey2010, gill2011modeling, serov2015analytical}.  Here we establish how the elaborate and irregular three-dimensional (3D) organisation of capillaries within terminal villi, the primary functional exchange units of the feto-placental circulation, contributes to solute exchange.

To maximize functional understanding from emerging 3D structural data requires an integrated mix of \emph{ex vivo} experiments \cite{Nye2018,Sibley_etal18} and computational modeling \cite{gill2011modeling,clark2015multiscale,mayo2016computational,mayo2016three,pearce2016image,RennieSled_etal17,bappoo2017viscosity}.  Previous studies have demonstrated how transport of highly diffusive solutes in capillaries with small diffusion distances is flow-limited (determined by the flow rate of fetal {or maternal} blood), whereas transport of slowly diffusing solutes in capillaries with a thick villous membrane is diffusion-limited.
While research has begun to shed light on the relationship between these transport regimes in the human placenta \cite{
Faber95,mayo2016three,pearce2016image}, the latest imaging data allow for a significantly more comprehensive characterization of the dominant geometric features and physical processes that govern solute transport in the placental microvasculature.   Quantifying such structure-function relationships is essential in building well-grounded multiscale models for whole-organ function of the human placenta and other complex vascular systems \cite{Moore15, peyrounette2018multiscale, Siggers_etal10, clark2015multiscale, Hunter14}. 

In this study, we use an integrative approach. We combine image analysis and 3D simulations with a discrete network model and asymptotic analysis to examine the dependence of solute transport on the geometrical arrangement of capillaries within terminal villi.
The properties of these functional exchange units are quantified and encapsulated in a theory of feto-placental transport (formulated \chg{as an} algebraic relationship) that links the complex 3D structure of fetal microvascular networks to their solute exchange capacity, providing a valuable building block for organ-level models.  We test the reduced scaling relationship against image-based computations and find that it applies both at the level of the whole network and within individual capillaries (subject to variations due to countercurrent effects), readily incorporating non-Newtonian effects of whole blood.  Our results suggest that \chgchg{an archetypal} physical scaling of feto-placental solute transport based on geometrical statistics provides a viable method for linking placental structure and function. Furthermore, \chg{our} developed and cross-validated framework offers significant savings in computational costs associated with image-based models of complex biological structures and could be applicable to other systems in which transport occurs via advection and diffusion in disordered microscale networks.

\begin{table}
\centering
\begin{tabular}{|c|c|c|c|c|}
\hline 
Specimen & 1 & 2 & 3 & 4\tabularnewline
\hline 
\hline 
{$\mathcal{R}/\eta\:\times10^{7}\,[\!\units{mm^{-3}}]$} & $7.4$ & $3.5$ & $27.9$ & $28.0$\tabularnewline
\hline 
$\mathcal{L}\:[\text{mm}]$ & $8.2$ & $11.4$ & $15.4$ & $17.9$\tabularnewline
\hline 
$L_\text{c}\:[\text{mm}]$ & $2.2$ & $1.8$ & $2.2$ & $2.3$\tabularnewline
\hline 
$\mathcal{L}/L_\text{c}$ & $3.7$ & $6.5$ & $7.0$ & $7.7$\tabularnewline
\hline 
\end{tabular}
\caption{Geometric parameters for network Specimens 1-4. The viscous resistance $\mathcal{R}$ scaled by the blood viscosity $\eta$ and \chgrev{the diffusive lengthscale specific to the villus} (integrated ratio of exchange area over exchange distance) $\mathcal{L}$ are determined computationally \chgrev{(see S.I., Sec.~2)}; the total centerline length $L_\mathrm{c}$ is determined through a skeletonization algorithm of the capillary network which provides vessel centerlines. 
}
\label{tab:geometric-parameters}
\end{table}

\section*{Results}

\begin{figure*} 
\centering
\includegraphics[width=1\textwidth]{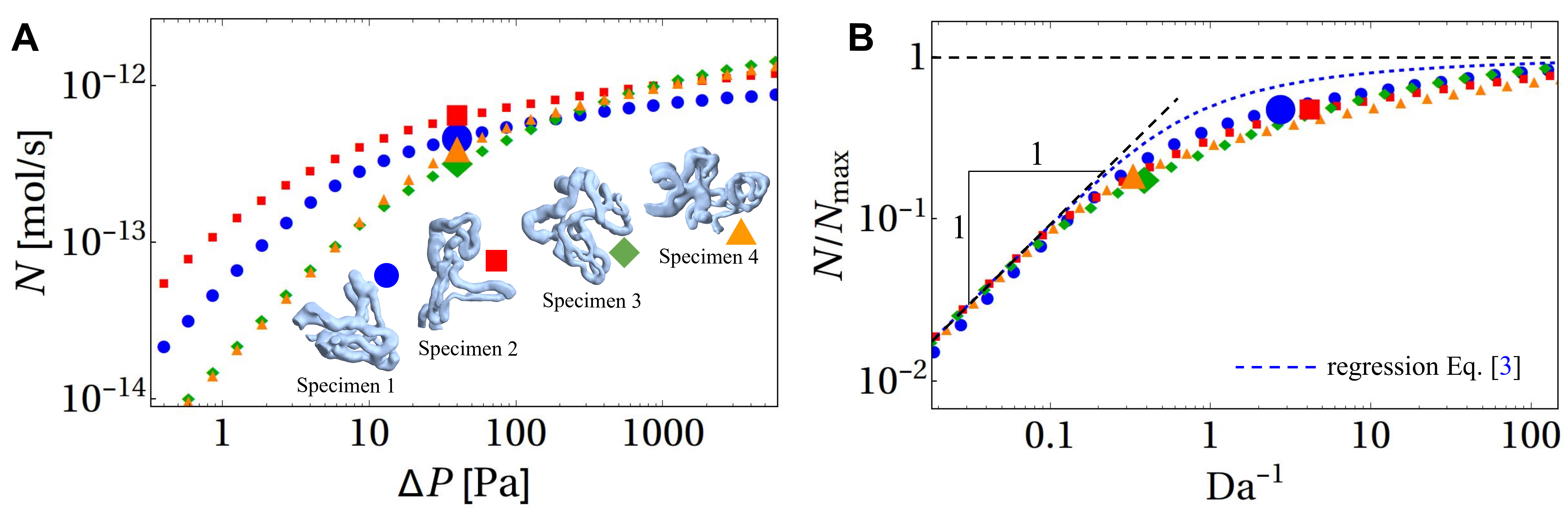}
\caption{Computational data (A) \chgchg{show appreciable} collapse when plotted using suitable dimensionless variables (B). \textbf{A} Computed solute flux $N$ in four segmented villus networks (Specimens 1-4) plotted against the pressure drop $\Delta P$ driving flow through each network.  \textbf{B} The same data presented in terms of the inverse Damk\"ohler number (see Eq.~\eqref{eq:Da}) and solute flux scaled on each specimen's diffusion-limited upper bound $N_{\text{max}}$.  $\mathrm{Da}^{-1}$ is proportional to the pressure drop $\Delta P$ driving flow through the network.  Predicted fluxes for each specimen (small colored symbols) \chgchg{collapse towards a common} relationship. Dashed lines show the approximation Eq.~\eqref{eq:regression} and its asymptotes. For Specimen 1, the largest deviation between the approximation Eq.~\eqref{eq:regression} and the computational result is $24\%$. The large symbols in A and B compare fluxes in each specimen evaluated at a fixed inlet-outlet pressure drop $\Delta P=40\units{Pa}$. We consider this value of $\Delta P$ physiological as it leads to shear stresses in Specimen 1 below approximately 1.2\,Pa, which we identify in S.I., Sec.~2 to be a physiological shear stress value.}
\label{fig:collapse}
\end{figure*}

\subsection*{Theory of solute transport in feto-placental networks}

The terminal villus shown in Fig.~1C is one of four samples we analyzed obtained by confocal laser scanning microscopy (from \cite{mayo2016computational,mayo2016three}).  Even within a single villus, there is significant variation in capillary diameters and \chgrev{exchange} distances \chgrev{between the capillary and villous surfaces} (see Fig.~S1 in the Supplementary Information, S.I.).   Image segmentation (S.I., Sec.~1) reveals the domains occupied by blood vessels ($\Omega_\text{b}$) and villous tissue ($\Omega_\text{t}$), as well as the bounding syncytiotrophoblast, which provides an interface $\Gamma_{\text{vil}}$ with maternal blood.  For each sample, identifying likely inlet and outlet vessels, we computed Stokes flow through the vessel network in $\Omega_\text{b}$ (non-Newtonian features of blood rheology are addressed below) under an imposed pressure drop $\Delta P$ to determine the network resistance $\mathcal{R}$ (Table~\ref{tab:geometric-parameters}).  Solute transport was computed using a linear advection-diffusion equation in $\Omega_\text{b}$ (modifying the advection term by a factor $B$ to account for facilitation of solute transport by the red blood cells), coupled to a diffusion equation in $\Omega_\text{t}$: solute concentrations differing by a value $\Delta c$ were prescribed on $\Gamma_{\text{vil}}$ and the inlet to $\Omega_\text{b}$ and the net flux $N$ of solute out of $\Omega_\text{b}$ was evaluated.  Solute uptake by tissue is not accounted for in this study.  Full details of the simulations are provided in Sec.~2 of the S.I.

For each of the four specimens (illustrated in Fig.~S1A), the computed net solute flux (evaluated using parameter values appropriate for oxygen) rises monotonically with the imposed pressure drop (Fig.~\ref{fig:collapse}A).  We wish to establish how the differing structures of each network lead to differences in the relationship between $N$ and $\Delta P$.  This understanding is facilitated by identifying the relevant dimensionless parameters and variables describing transport in this functional tissue unit \cite{Hunter14}.

Flow-limited transport arises when $\Delta P$ is sufficiently weak for solute to be fully saturated in fetal blood before it leaves the vessel network.  In this case $N$ is determined by the \chgrev{flow rate} through the outlet $(\Delta P/\mathcal{R})$ as $N=\Delta c\, B \Delta P/ \mathcal{R}$ (where $B$ models facilitated transport).   In contrast, an upper bound on $N$ arises when the transport is diffusion-limited, with flow being sufficiently rapid to impose the fixed concentration difference $\Delta c$ between $\Gamma_{\text{vil}}$ and the boundary $\Gamma_{\text{cap}}$ (the capillary endothelium separating $\Omega_\text{b}$ from $\Omega_\text{t}$).  In this case $N=N_{\text{max}}\equiv D_\text{t}\, \Delta c\, \mathcal{L}$, where $\mathcal{L}$ is a lengthscale specific to the villus and $D_\text{t}$ is the solute diffusivity in tissue \cite{jensenchernyavsky2018}.  ($\mathcal{L}$ can be evaluated by solving Laplace's equation $\nabla^2 c=0$ in $\Omega_\text{t}$ with $c=0$ on $\Gamma_{\text{cap}}$ \chgchg{and $c=\Delta c$ on $\Gamma_{\text{vil}}$, and} integrating the normal gradient of $c$ over either $\Gamma_{\text{cap}}$ or $\Gamma_{\text{vil}}$; see S.I. Sec.~2). 
\chgchg{We can compare the diffusive capacity per unit concentration across the villous tissue, $D_t\mathcal{L}$,  with a dimensionally-equivalent measure of diffusive capacity along vessels using the dimensionless parameter}
\begin{equation}
\mu=\frac{D_\text{t}\,\mathcal{L}}{D_\text{p}\,L_\text{c}},
\label{eq:mu}
\end{equation}
where $D_\text{p}$ is the solute diffusivity in blood plasma \chgrev{and $L_\text{c}$ is a measure of vessel length in the villus. Taking $L_\text{c}$ as the total centreline length of capillaries within the network,} it is notable that the ratio $\mathcal{L}/L_\text{c}$ shows \chgchg{only modest} variation between specimens (Table~\ref{tab:geometric-parameters}), despite significant variability in network structure (Fig.~\ref{fig:collapse}A, insets).

The ratio of fluxes in the diffusion- and flow-limited states defines a dimensionless Damk\"{o}hler number
\begin{equation}
\mathrm{Da} =\frac{D_\text{t}\,\mathcal{L}\,\mathcal{R}}{B\,\Delta P},
\label{eq:Da}
\end{equation}
which also has an interpretation as a ratio of a timescale for advection within the vessel network to a diffusive timescale through the tissue. The parameters  $\mu$ and $\mathrm{Da}$ are convenient for  {characterizing} solute exchange in a terminal villus \chg{\cite{jensenchernyavsky2018}}, as illustrated for a single vessel in Sec.~3 of the S.I. 

For each villus sample, we computed three geometric determinants of transport, $L_c$, $\mathcal{L}$ and $\mathcal{R}/\eta$ (see Table~1; \chg{in simulations we} used uniform blood viscosity $\eta=2\times 10^{-3}\units{Pa\!\cdot\!s}$).   The $\mathcal{R}$ and $\mathcal{L}$ values are larger for Specimens 3 and 4 than for Specimens 1 and 2, likely because the latter were fixed at approximately three times higher \chg{fetal} perfusion pressure \chg{(see Materials and Methods). It is notable that differences revealed by these global measures are not obviously captured by simpler summary statistics such as average capillary radii (Fig.~S1)}.
We then replotted the relation between net flux $N$ and pressure drop $\Delta P$ in terms of $N/N_{\text{max}}$ (scaling flux on the diffusion-limited upper bound) and $\mathrm{Da}^{-1}$ (the natural dimensionless proxy for $\Delta P$).  These variables incorporate dependencies on the material parameters $B$, $D_\text{t}$ and $D_\text{p}$, which we report for different solutes in Table~S1. Despite substantial variation in network structure, the data \chgchg{collapse appreciably} (Fig.~\ref{fig:collapse}B), \chgchg{showing a common smooth transition} between flow-limited and diffusion-limited transport as $\mathrm{Da}^{-1}$ increases. The large symbols in Fig.~\ref{fig:collapse} \chgchg{show how, at a fixed physiological inlet-outlet pressure drop $\Delta P = 40\units{Pa}$ (S.I., Sec.~2), geometric differences in flow resistance between specimens lead to different inverse Damk\"ohler numbers $\mathrm{Da}^{-1}$ (Fig.~2B)}.

Extending a regression formula proposed previously \cite{pearce2016image,jensenchernyavsky2018}, we approximate \chg{the} relationship between $N$ and $\mathrm{Da}^{-1}$ (S.I., Sec.~3) using
\begin{equation}
N=\frac{N_{\text{max}}}{\text{Da}\left(1-e^{-\text{Da}}\right)^{-1}+\text{Da}_\mathrm{F}^{1/3}},
\label{eq:regression}
\end{equation}
which captures the simulated fluxes with a reasonable degree of accuracy (Fig.~\ref{fig:collapse}B).  Here the parameter $\mathrm{Da}_\text{F}=\mu^2\, \mathrm{Da}/\alpha_\text{c}^3$, where $\alpha_\text{c}\approx 5.5$, accounts for transport across concentration boundary-layers within capillaries \cite{pearce2016image}.  Setting this term to one side for a moment, the remaining terms provide a smooth transition between flow-limited transport \chg{(}$N\approx N_{\text{max}}/\mathrm{Da}$ when $\mathrm{Da}^{-1}\ll 1$\chg{)} and diffusion-limited transport \chg{(}$N\approx N_{\text{max}}$ when $\mathrm{Da}^{-1}\gg 1$\chg{,} Fig.~\ref{fig:collapse}B).  \chg{Despite substantial variation in network structure, the data \chgchg{collapse towards a common relationship} (Fig.~\ref{fig:collapse}B) in the flow-limited ($\mathrm{Da}^{-1}\ll 1$) and diffusion-limited ($\mathrm{Da}^{-1}\gg 1$) regimes, \chgchg{while \chgrev{showing similar} qualitative behavior in the transitional region for $\mathrm{Da}=O(1)$.}}

\begin{figure}[t] 
\centering
\includegraphics[width=0.49\textwidth]{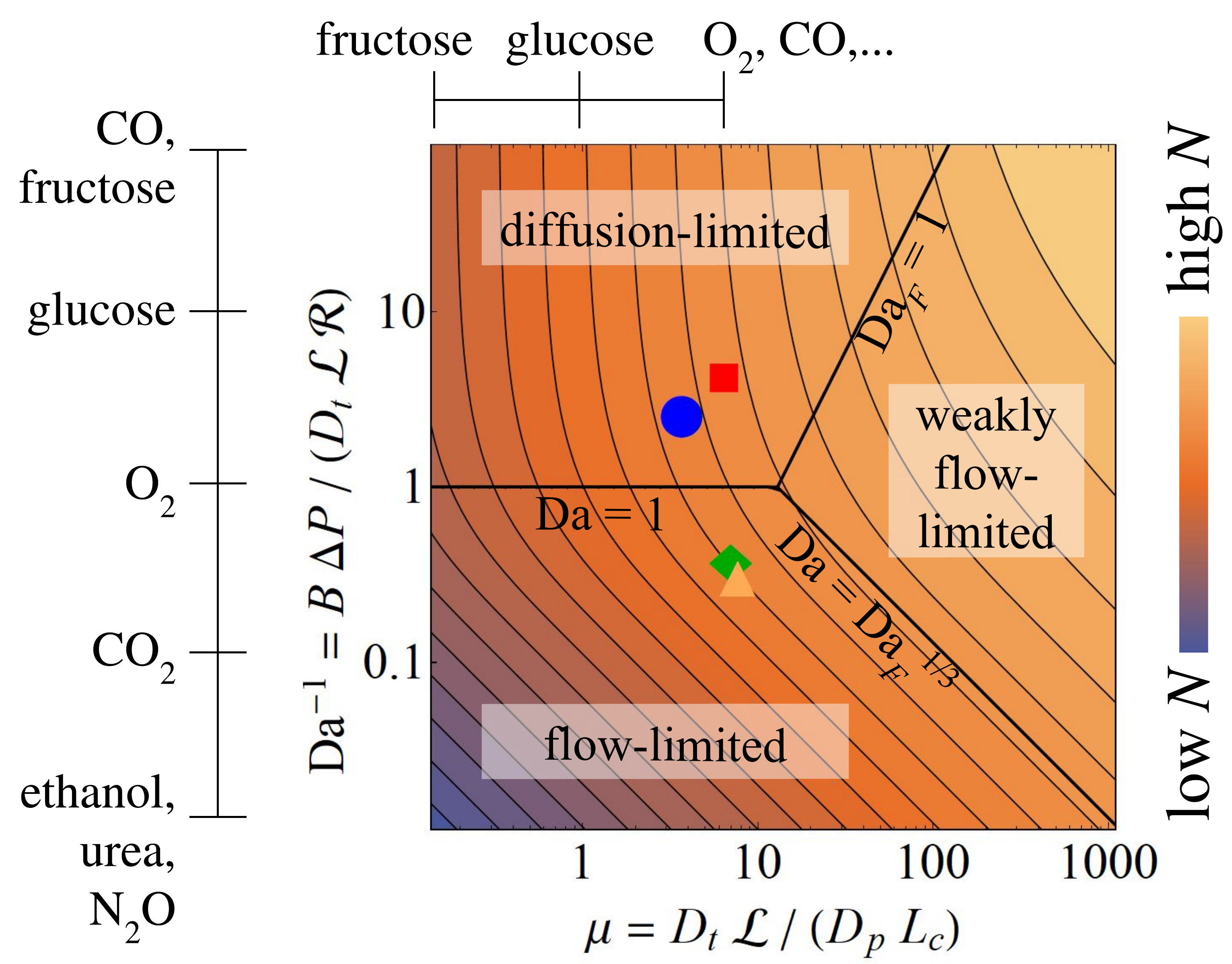}
\caption{{A diagram summarizing transport regimes in the parameter space spanned by $\mu$ (see Eq.~\eqref{eq:mu}), measuring the tissue's capacity for diffusive transport relative to diffusion in the vessel network, and $\mathrm{Da}^{-1}$ (see Eq.~\eqref{eq:Da}) which is proportional to flow.   Contours and background color indicate the network solute flux $N$ (see Eq.~\eqref{eq:regression}), evaluated for fixed $\Delta c$ and $L_\text{c}$.   The  diffusion-limited regime ($\mathrm{Da}^{-1}\gg \max(1,\mu^2)$), for which $N\approx N_{\text{max}}$, and  two flow-limited regimes are indicated.  In the strongly flow-limited state ($\mathrm{Da}^{-1} \ll \min(1,\mu^{-1})$), flux is proportional to flow ($N\approx N_{\text{max}} \mathrm{Da}^{-1}$), corresponding to an asymptote shown in Fig.~\ref{fig:collapse}{B}.  In the weakly flow-limited state ($\mu^{-1}\ll \mathrm{Da}^{-1}\ll \mu$), concentration boundary-layers arise within capillaries and $N\approx N_{\text{max}} \mathrm{Da}^{-1/3}/\mu^{2/3}$. The large colored symbols correspond to those in Fig.~\ref{fig:collapse}, placing oxygen transport in Specimens 1--4 well outside the weakly flow-limited regime, spanning the interface of strongly flow-limited and diffusion-limited regimes.  Vertical and horizontal bars outside the figure indicate the relative $\mu$ and $\mathrm{Da}^{-1}$ values of a variety of solutes {with respect to oxygen}, based on data in Table~S1. { The upper limits of the ranges of Table~S1 are shown. For instance, $\mathrm{Da}^{-1}$ of glucose is approximately ten times higher compared to oxygen, and $\mu$ of glucose is approximately ten times lower.}
}}
\label{fig:region-diagram}
\end{figure}

This transition is illustrated on the left-hand side of the regime diagram in Fig.~\ref{fig:region-diagram}.  The symbols show how, {imposing a physiological inlet-outlet pressure drop $\Delta P=40\units{Pa}$} across all four specimens, oxygen fluxes span the transition between flow- and diffusion-limited states.  Eq.~\eqref{eq:regression} suggests that, for villi and solutes having sufficiently large $\mu$ (\hbox{i.e.} rapid transmural diffusive transport), boundary-layer effects may emerge \cite{pearce2016image}, introducing an intermediate weakly-flow-limited state for intermediate values of $\mathrm{Da}$.  However, our simulations demonstrate that, for oxygen transport in the four samples investigated, $\mu$ is sufficiently small for this not to be relevant under normal conditions. \chgchg{Fig.~\ref{fig:region-diagram} also shows that, between different specimens, $\mathrm{Da}$ spreads over more than an order of magnitude, \chgchg{ for a given $\Delta P$, reflecting differing flow resistances among villi.} In contrast, the ratio $\mathcal{L}/L_\text{c}$, and hence the parameter $\mu$ (Eq.~\eqref{eq:mu}), varies by approximately a factor of 2, as revealed by Table~\ref{tab:geometric-parameters}.}

We can extend this analysis to a variety of small and mobile solutes using the data in Table~S1, which summarizes estimated effective advection-enhancement factors $B$, plasma diffusivities $D_\text{p}$ and tissue diffusivities $D_\text{t}$. From these we compute \chg{inverse Damk\"ohler} numbers $\mathrm{Da}_{\text{rel}}^{-1}$ relative to the value for oxygen. Taking oxygen transport as a reference, we identify strongly diffusion-limited solutes, such as mannitol, fructose or carbon monoxide (for which $\mathrm{Da}_\text{rel}^{-1} \gg 1$) as well as strongly flow-limited solutes, including certain anaesthetic gases (e.g. nitrous oxide), urea and ethanol (for which $\mathrm{Da}_\text{rel}^{-1} \ll 1$). It is noteworthy that the transport regime in which a solute lies (see Fig.~\ref{fig:region-diagram}) is affected by inverse \chg{Damk\"ohler} number through the \chg{ratio} $B/D_\text{t}$, and affected by the diffusive capacity \chg{ratio} $\mu$ through the ratio $D_\text{t} / D_\text{p}$. As Table~S1 shows, for a fixed geometry $\mathrm{Da}$ has a much wider spread than $\mu$ through large variability of $B$, which ranges over four orders of magnitude. 
However, the maximum achievable flux $N_\text{max}$ is proportional to $D_\text{t}$ alone, and therefore $N_\text{max}$ values for oxygen and CO are predicted to be almost twice those of ethanol and caffeine for the same transmural concentration difference (Table~S1).

\rem{ 
} 


\begin{figure*}[t!] 
\centering
\includegraphics[width=1\textwidth]{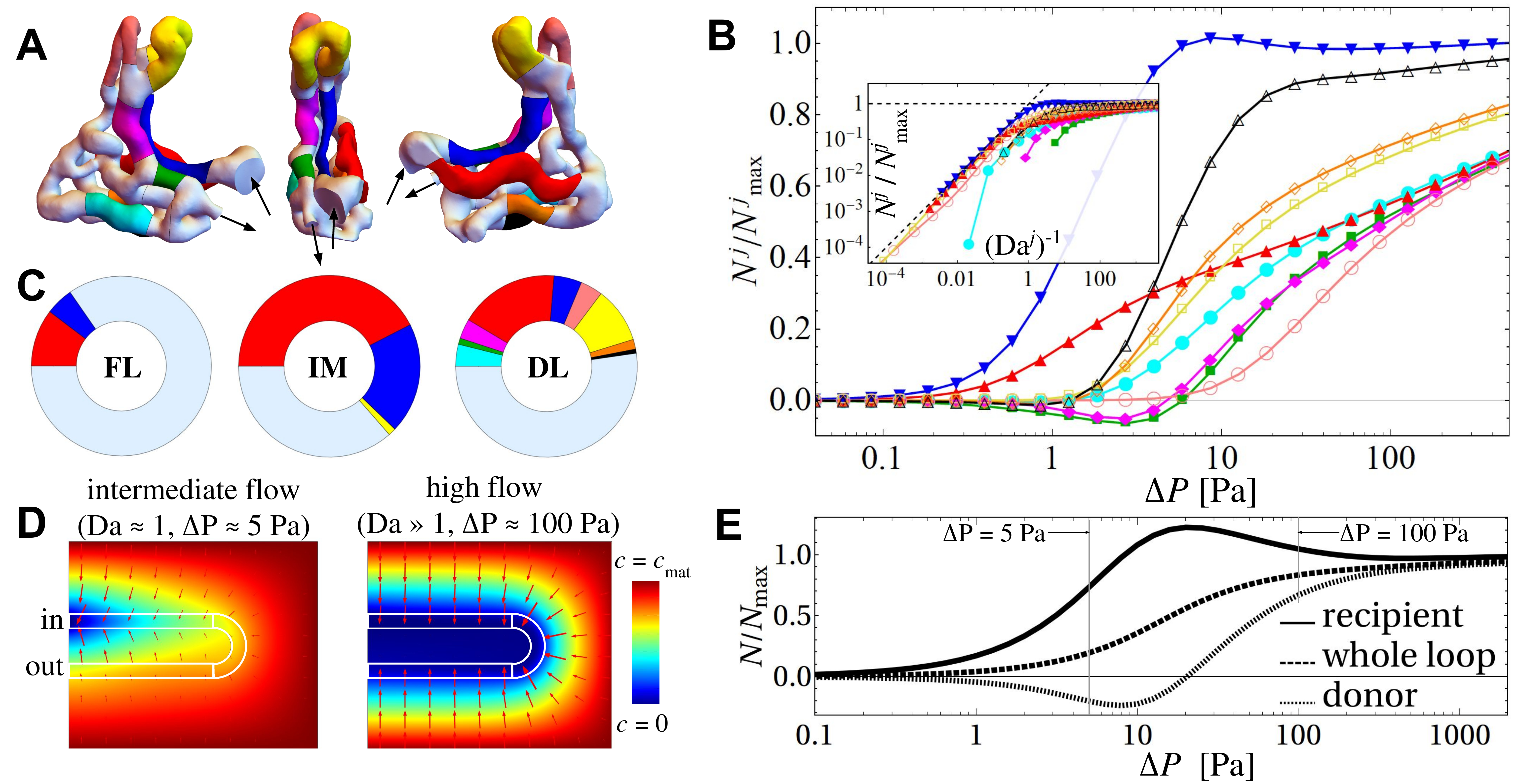}
\caption{
Solute exchange heterogeneity at the level of individual capillaries.  \textbf{A} The nine longest capillaries of Specimen 1 are highlighted in color; the rest of the network is shown in light blue. Arrows indicate inlet and outlet in 3 projections of the network. The blue capillary near the inlet neighbors green and magenta capillaries near the outlet; likewise red (near inlet) neighbors orange and black (near outlet). \textbf{B} The scaled net uptake of vessel $j$, $N^j/N^j_\text{max}$, as a function of the pressure drop $\Delta P$ across the whole network 
exhibits non-monotonicity in some cases, due to a donor-recipient mechanism explored in panels D \& E. The inset shows a log-log plot of the same data as a function of $(\mathrm{Da}^j)^{-1}$, highlighting a collapse of the data similarly to the whole network (Fig.~\ref{fig:collapse}B), with the exception of donor capillaries for which $N$ becomes negative (truncated curves). \textbf{C} Relative contributions of different capillaries to net uptake of the entire network. The inlet-outlet pressure drop in the flow-limited (FL) regime is {$\Delta P=0.04 \units{Pa}$}, in the intermediate (IM) regime {$\Delta P=1.26 \units{Pa}$}, and in the diffusion-limited (DL) regime {$\Delta P=186 \units{Pa}$}. \textbf{D} Simplified capillary loop model system of donor-recipient mechanism, from a computation in two spatial dimensions. Red arrows illustrate \chgrev{directions of} diffusive flux in the surrounding tissue; capillary boundaries are white lines. At intermediate pressure drops, a counter-current effect extracts solute from the bottom capillary (acting as a donor) into the top capillary (recipient). The net fluxes of the inlet \textit{recipient} and outlet \textit{donor} capillaries as a function of pressure drop (\textbf{E}) show the same characteristic behavior as demonstrated in panel B: at intermediate pressure drops the donor(s) switch sign whereas the recipient surpasses its carrying capacity $N_\text{max}$, but this effect is integrated out at the level of the whole system (\textit{whole loop} in E).}
\label{fig:individual-capillaries}
\end{figure*}

\subsection*{Network heterogeneity}
{To understand spatial variations in solute transfer within capillary networks, we now focus} on solute exchange at the level of individual capillaries. For the nine longest capillaries of Specimen 1 (highlighted in Fig.~\ref{fig:individual-capillaries}A and labelled by $j$), we evaluted the scaled net uptake, $N^j/N^j_\text{max}$, as a function of the pressure drop $\Delta P$ across the whole network (see the log-linear plot in Fig.~\ref{fig:individual-capillaries}B). The scaled net uptake exhibits heterogeneity across the sample of vessels, including non-monotonicity in some cases. In particular, uptake in the blue capillary surpasses its carrying capacity $N_\text{max}$ at intermediate $\Delta P$. Conversely, transport in the neighboring magenta and green capillaries switches sign around the same intermediate pressure-drop regime, suggesting a change in their role from donors of oxygen at low $\Delta P$ to recipients at high $\Delta P$ (via a mechanism explored in Fig.~\ref{fig:individual-capillaries}D \& E). The inset shows a log-log plot of the same data as a function of $(\mathrm{Da}^j)^{-1}$, highlighting a collapse of the data similarly to the whole network (Fig.~\ref{fig:collapse}B), with the exception of donor capillaries for which $N$ becomes negative (truncated curves). 

To illustrate the donor-recipient mechanism, we consider a simplified model system in Fig.~\ref{fig:individual-capillaries}D. A capillary loop, embedded in a box of villous tissue, carries solute from the inlet (top) to the outlet (bottom) capillary.   At intermediate pressure drops a counter-current effect extracts solute from the outlet capillary (acting as a donor) into the inlet capillary (the recipient). The net flux of the top and bottom capillaries as a function of pressure drop (Fig.~\ref{fig:individual-capillaries}E) shows the same characteristic behavior as demonstrated in Fig.~\ref{fig:individual-capillaries}B: at intermediate $\Delta P$ the donor flux switches sign whereas the recipient surpasses its carrying capacity $N_\text{max}$. At the level of the entire loop, however, the net uptake $N$ neither surpasses the carrying capacity $N_\text{max}$, nor does it become negative. Similarly, the heterogeneity seen in individual vessels of the Specimen 1 capillary network (Fig.~\ref{fig:individual-capillaries}B) is integrated out at the level of the entire network (Fig.~\ref{fig:collapse}A).

\begin{figure*}[t] 
\centering
\includegraphics[width=1\textwidth]{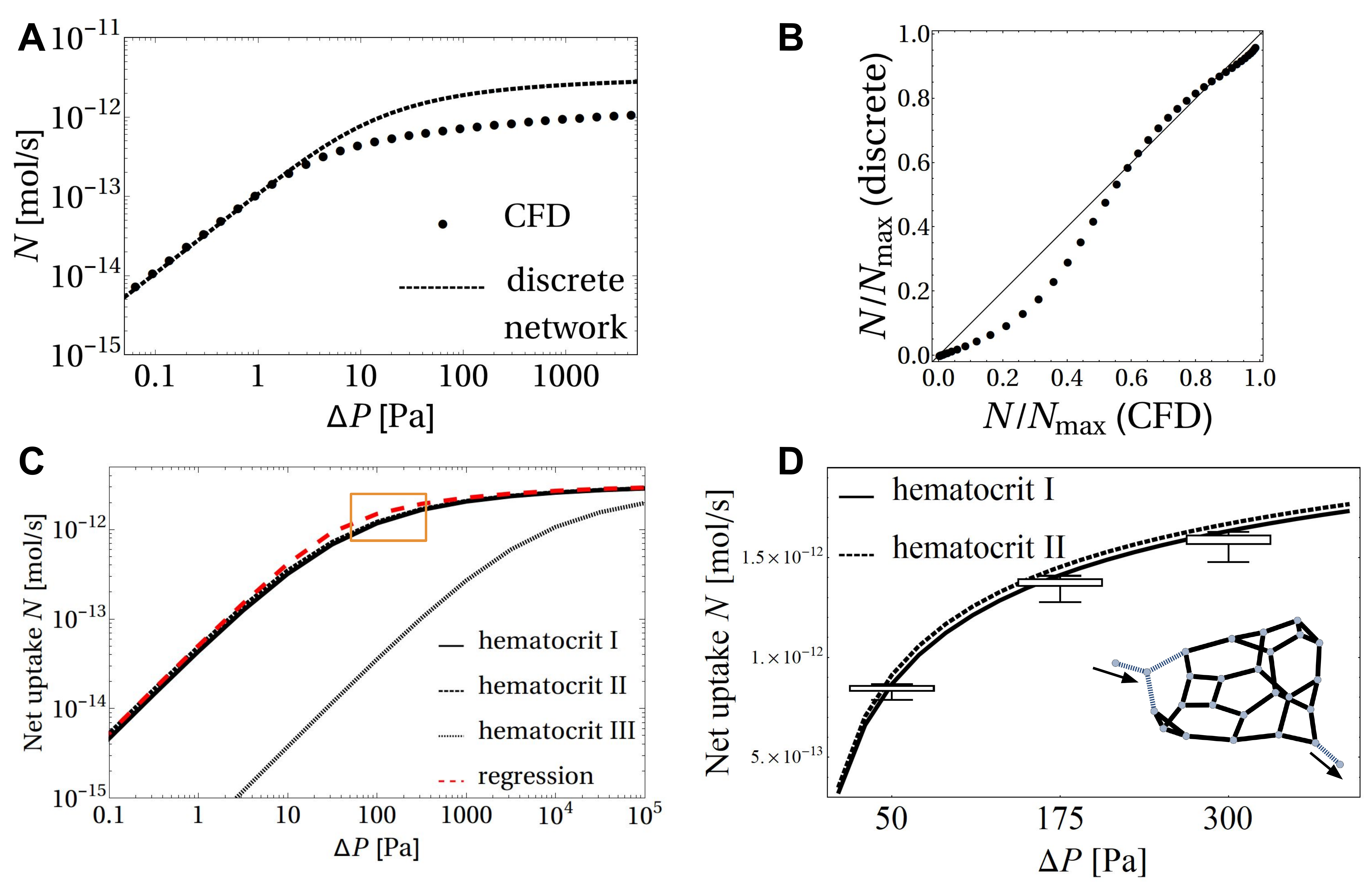}
\caption{A comparison \chgrev{between} the discrete network vs.\ CFD models of oxygen transfer in Specimen 1 (topology shown as an inset to \textbf{D}). \textbf{A} compares the solute flux $N$ versus network pressure drop $\Delta P$ as predicted by the computational model (\chgrev{S.I.,} Sec.~\ref{sec:computational-model}) and the discrete network model (\chgrev{S.I.,} Sec.~\ref{sec:discrete-network-model}).  \textbf{B} shows the same data when rescaled by relevant values of $N_{\text{max}}$. \textbf{C}~Dependence of the discrete network-predicted oxygen net transfer rate $N$ on hematocrit distribution.  The oxygen transfer rate for varying $\Delta P$ for the entire network is predicted assuming uniform hematocrit and facilitated transport ($B=141$, \emph{hematocrit I}, solid line), spatially variable hematocrit ($B=B(H)$, \emph{hematocrit II}, dashed line) and uniform hematocrit but without facilitated transport ($B=1$, \emph{hematocrit III}, thin-dashed line). \chgrev{The regression equation Eq.~\eqref{eq:regression} applied to the entire discrete network is shown as the red dashed line (see S.I., Sec.~\ref{sec:discrete-network-model}).} \textbf{D} Sensitivity of net oxygen flux to removal of individual vessels. The solid curves replicate those within the orange box in C.  For three different pressure drops ($\Delta P=50 \units{Pa}, 175 \units{Pa}$ and $300 \units{Pa}$), we calculated 33 values of $N$ with each of the 33 black capillaries (inset) removed individually. The resulting distribution for the non-uniform hematocrit model is shown with box plots, demonstrating that the network is robust with respect to the occlusion of individual capillaries.}
\label{fig:discrete-model}
\end{figure*}

Clarification of the donor-recipient mechanism adds to our understanding of the  contributions of individual vessels to the overall solute transfer of the capillary network, shown in Fig.~\ref{fig:individual-capillaries}C. \chgGreen{For a low inlet-outlet pressure drop, the network is situated in} the flow-limited (FL) regime, \chgGreen{where} practically all uptake is reduced to a narrow region near the inlet. Among the nine colored capillaries, only the blue and red one are close to the inlet, adding a small contribution each. In the intermediate (IM) regime, the donor-recipient effect peaks, favoring the blue recipient capillary at the expense of the neighboring green and magenta donors from which solute is extracted (and, to a lesser extent, the red at the expense of orange and black). In the diffusion-limited (DL) regime, capillaries at the periphery of the network, in proximity to a large portion of the surrounding villous surface (particularly the red and yellow capillaries), add the greatest contributions to transport. \chg{Figure~\ref{fig:individual-capillaries}C therefore illustrates how different vessels contribute to transport as the network moves from a flow-limited to a diffusion-limited state across Fig.~\ref{fig:region-diagram}.} 

The computational results underlying Figs~\ref{fig:collapse}--\ref{fig:individual-capillaries} are based on a Newtonian transport model \chgrev{with uniform hematocrit,}
evaluated using \chg{3D finite-element} simulations. In order to assess the non-Newtonian effects of hematocrit on solute transport, we developed a discrete network model \chgrev{(see S.I., Sec.~4)} which relies on the well-established semi-empirical Pries--Secomb model \cite{pries1990blood}, implemented in a reduced representation of each network in which each capillary is treated as a discrete component (S.I., Sec.~4).
\chgrev{Fig.~\ref{fig:discrete-model}A,B compares predictions of the reduced (discrete network) model to the full (computational fluid dynamics, CFD) model for uniform hematorcrit and blood viscosity. Although the discrete network model captures the scaling relationship between the uptake flux $N$ and pressure drop $\Delta P$ (Fig.~\ref{fig:discrete-model}C) and shows a good overall agreement with the CFD (Fig.~\ref{fig:discrete-model}A,B), the discrete network model overestimates $N$ at large $\Delta P$ and underestimates $N$ at small $\Delta P$ (see Discussion for further context).} 
\chgrev{Fig.~\ref{fig:discrete-model}C compares the net oxygen transfer, assuming either uniform hematocrit and blood viscosity (\emph{hematocrit I}, where $H=0.48$, $\eta=2\times 10^{-3}\units{Pa\!\cdot\!s}$, $B=141$) or spatially variable hematocrit and nonlinear Pries--Secomb blood rheology (\emph{hematocrit II}, where the effective viscosity $\eta(H)$ and solute carrying capacity $B(H)$ vary across the network).}
While the F{\aa}hr{\ae}us--Lindqvist effect can be expected to lower the net resistance of flow through the network, enhancing $N$ for a given $\Delta P$, the hematocrit reduction in smaller vessels due to plasma skimming reduces their oxygen carrying capacity.  Fig.~\chgrev{\ref{fig:discrete-model}}C shows how, for Specimen 1, the two effects are predicted to counteract, leading to modest net impact on overall oxygen transport, supporting the use of the Newtonian model \chgrev{and, furthermore, preserving the predictive power of the scaling relationship \eqref{eq:regression} in the discrete network model}. \chgrev{However, the impact of solute carrying capacity is significant (Fig.~\ref{fig:discrete-model}C): setting $B=1$ (\emph{hematocrit III}) to eliminate the effect of solute binding to hemoglobin substantially reduces $N$ compared to \emph{hematocrit I} and \emph{hematocrit II}, particularly under flow-limited conditions.}

We also used the discrete network model to probe the sensitivity of oxygen transport to removal (or temporary blockage) of individual vessels. 
We calculated distributions of network oxygen transfer $N$ when individual capillaries of Specimen 1 are removed from the network (excluding those very close to the inlet). Removal of a single vessel reduces the overall network transfer by no more than 10\% (see Fig.~\chgrev{\ref{fig:discrete-model}D}), demonstrating the robustness of the network to the occlusion of individual capillaries.

\section*{Discussion}
This study demonstrates how, despite highly variable network geometries, solute transfer between maternal and fetal circulations in a terminal villus can be {characterized} effectively using two dimensionless parameters (the diffusive capacity \chg{ratio} $\mu$ and the Damk\"ohler number $\text{Da}$, see Eqs~\eqref{eq:mu}, \eqref{eq:Da}), which in turn depend on three geometry-dependent dimensional quantities (the total centerline length of capillaries within a network $L_\text{c}$, the diffusive lengthscale $\mathcal{L}$ relating capillary and villus geometry, and the network flow resistance $\mathcal{R}$).  These can be extracted from microscopy images via standard tools (finite-element analysis and image skeletonization) and provide a computational generalization for disordered tissues of the classical Krogh cylinder approach.  These variables reveal scaling relationships that hold both at the network and capillary levels: the appropriate choices of $\mu$ and $\text{Da}$ lead to a \chgchg{near-}collapse of transport behavior across multiple terminal villi (Fig.~\ref{fig:collapse}B), as well as for individual capillaries within a villus network (Fig.~\ref{fig:individual-capillaries}B).  Furthermore, the algebraic approximation Eq.~\eqref{eq:regression} compactly summarizes the transport capacity of a villus.  Its transparent dependence on physical parameters gives immediate insights into the physical and geometric determinants of solute transport, and its economy makes it attractive as a component in future multiscale models of placental function. 

The model readily describes transfer of a variety of passively transported solutes.  Varying diffusion coefficients and the binding capacity to hemoglobin influences $\mu$ and $\mathrm{Da}$, revealing solutes that are predominantly flow- or diffusion-limited (Table~S1).  The wide spread of parameter values illustrated in Fig.~\ref{fig:region-diagram} ($\mathrm{Da}$ spans four orders of magnitude) emphasizes how flow- and diffusion-limited transport are likely to occur concurrently in a single villus for different solutes \cite{Faber95}.
It remains to be seen whether the relatively modest variation in $\mu$ compared to $\mathrm{Da}$ (Fig.~\ref{fig:region-diagram}) for oxygen and other mobile solutes indicates a possible robust design feature of feto-placental microvasculature, which could be mediated  in the developing placenta by the dynamic balance of angiogenesis and vascular pruning~\cite{Benjamin98}.

\chgrev{A one-dimensional discrete network model (Fig.~\ref{fig:discrete-model}) offers a level of detail intermediate between the full 3D computational and algebraic regression (Eq.~\eqref{eq:regression}) approximations, enabling the analysis of feto-placental transport performance at minimal computational and image-processing costs. The discrete network model matches the predictions of the computational model in the physiological range of capillary pressure drops (Fig.~\ref{fig:discrete-model}A,B); however, it overestimates the uptake flux for fast flows (in the diffusion-limited transport regime) due to its neglect of diffusive shielding, i.e. spatial interaction between neighbouring capillaries (see e.g. Fig.~\ref{fig:individual-capillaries}D). The diffusive shielding is captured in 3D via $\mathcal{L}$ by integrating over the whole tissue domain, extending prior studies in 2D~\cite{gill2011modeling}. Likewise, the discrete model overestimates the network flow resistance and thus underestimates the uptake flux at small pressure drops (in the flow-limited transport regime), due to the strong (fourth-power) sensitivity of resistance on capillary radii, which are more accurately captured by the integral resistance $\mathcal{R}$ of the 3D computational model.}

The present model exploits emerging anatomical data for terminal villi but has some significant limitations. Our calculations over a discrete vessel network using the Pries--Secomb model \cite{pries1990blood}, which {characterizes} hematocrit distributions in individual cylindrical vessels, suggest that the effect of non-Newtonian blood rheology on oxygen transport is modest (Fig.~\chgrev{\ref{fig:discrete-model}}C), and that the network itself is robust to occlusions of individual vessels (Fig.~\chgrev{\ref{fig:discrete-model}}D), {which may occur} transiently due \chg{(for example)} to red blood cells lingering at network bifurcations \cite{Bagchi2017}.   These predictions await confirmation through more detailed theoretical studies that {describe} blood rheology in complex geometrical domains, and suitable experimental observations.  We have not accounted for uptake of solutes by the placental tissue itself, which \chg{will} be a significant feature for solutes such as oxygen \chg{(and which could shift the transport into a more flow-limited regime)}; the predicted fluxes must therefore be treated as upper bounds until future studies address this feature in more detail.  We have also encountered a common problem in simulating flows through microvascular networks, namely in reliably identifying inlet and outlet vessels.  This choice influences vessels that may serve as donors or recipients when countercurrent effects arise in the flow-limited regime (Fig.~\ref{fig:individual-capillaries}); however the choice has negligible impact on net transport in the diffusion-limited regime.  We have also over-simplified the supply of solute at the villus surface; this will be influenced by local features of the flow of maternal blood in the intervillous space. 
\chgrev{\chgrev{The model also assumes negligible interstitial flow in the villous tissue and does not account for transport of certain solutes via paracellular channels or energy-dependent membrane transporters \cite{jensenchernyavsky2018,Sibley_etal18}}.}
\chg{Finally, our model does not explicitly account for nonlinear oxygen--haemoglobin binding kinetics \chgchg{(the effects of which are evaluated in \cite{pearce2016image})} and the particulate nature of capillary blood flow that could result in subtle spatial oxygen gradients (e.g. see \cite{Hellums_etal95} \chgchg{for an extensive overview}). While our modelling framework provides a \chgchg{robust qualitative} description of transport in complex microvascular networks for a wide variety of solutes, it \chgchg{requires further quantitative refinement} in future studies.}

A key message of this study is that, despite the significant variability in the shapes of individual capillaries within a terminal villus, the overall capacity of the villus to transport passive solutes can be captured using three integrated quantities ($L_\text{c}$, $\mathcal{L}$ and $\mathcal{R}$) which to some extent average out \chg{intrinsic} variations.  It remains to be seen to what extent local features such as isolated `hot-spots' of transfer (where well-perfused capillaries lie very close to the villus surface, for example) might correlate with features of the external maternal flow, or the distribution of transporters in the villus membrane.  Such features may lead to non-trivial coupling between fetal and maternal flow distributions \cite{jensenchernyavsky2018}.  Once suitable imaging data become available, it will be of particular interest to \chgchg{explore both intra- and inter-placental variability and to} examine how pathologies that disrupt the structure of terminal villi impact on their function.

In summary, our analysis demonstrates {how a judicious choice} of dimensionless variables, incorporating relevant integral determinants of geometric microstructure, {reveals} robust relationships characterizing physiological function.  We anticipate that the framework we propose for assessing feto-placental solute transport performance can usefully be extended to other complex microvascular systems.

\section*{Materials and Methods}
\chg{The \chgchg{specimens} were taken from two different peripherial lobules of a normal human placenta delivered by Cesarean section at term, as reported previously \cite{mayo2016three}. The lobules were fixed at different feto-placental fixation pressures (Specimens 1 and 2 at $100\text{ mmHg}$, Specimens 3 and 4 at $30\text{ mmHg}$, see \cite{mayo2016three}), and  the samples within each lobule were randomly sampled.}

Full details of the image analysis, 3D flow and transport simulations, discrete network model and sensitivity analysis are provided in the Supplementary Information. 
\chg{All data needed to evaluate the conclusions in the paper are present in the paper and the Supplementary Information. Additional data are available from the authors upon request.}



\begin{acknowledgments}
The authors thank John Aplin, Paul Brownbill, Edward D. Johnstone and Rohan M. Lewis for helpful discussions. This work was supported by the MRC (MR/N011538/1) and EPSRC (EP/K037145/1) research grants, and  by the Centre for Trophoblast Research, University of Cambridge.
\end{acknowledgments}

\nolinenumbers

\bibliographystyle{naturemag}
\bibliography{network-model}


\onecolumngrid
\pagebreak
\section*{\large{Supplementary Information}}
\renewcommand \thefigure   {S\arabic{figure}}
\renewcommand \thetable    {S\arabic{table}}
\renewcommand \thesection  {S\arabic{section}}
\counterwithout{equation}{section}
\renewcommand \theequation {S\arabic{equation}}
\setcounter{equation}{0}
\setcounter{figure}{0}
\setcounter{table}{0}

This Supplement provides further details on the image analysis (Sec.~\ref{sec:imstat}), the governing equations and methodology used in 3D computations (Sec.~\ref{sec:computational-model}), the asymptotic model of transport in a single vessel that motivates the regression equation (Sec.~\ref{sec:single-capillary-regression}), the discrete network transport model (Sec.~\ref{sec:discrete-network-model}) and its use in assessing the impact of nonlinear blood rheology and network heterogeneity. 
The associated structural datasets and computational codes are available on request.

\section{Image analysis and network statistics}
\label{sec:imstat}

\begin{figure}[t!]
\centering
\includegraphics[width=1\textwidth]{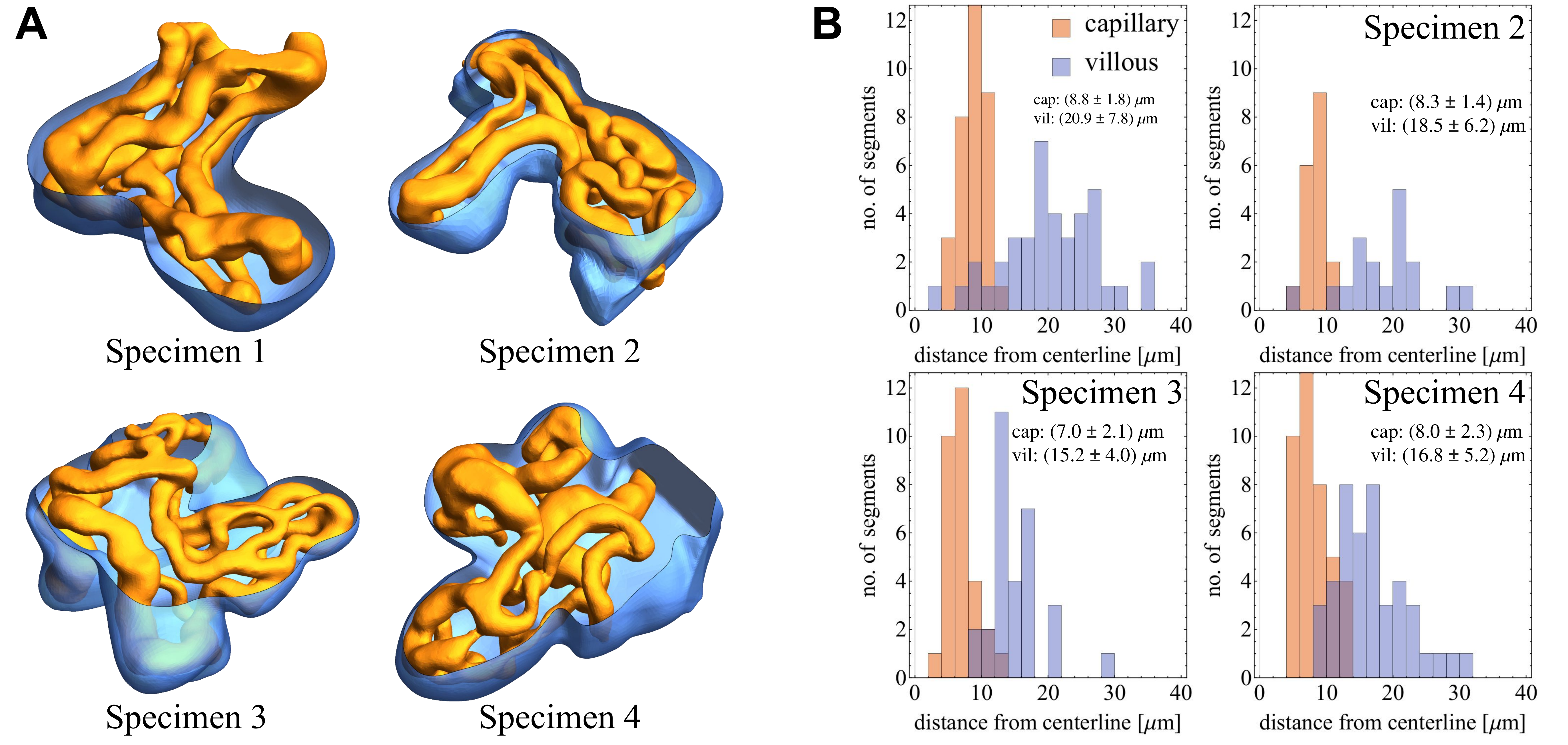}
\caption{\textbf{A} Four segmented terminal villi, showing capillary surface (rendered in yellow) and syncytiotrophoblast (blue). \textbf{B} \chg{Vessel-averaged minimal distance between centerline and $\Gamma_\mathrm{cap}$, as well as between centerline $\Gamma_\mathrm{vil}$. The vessel-averaging consists of discretizing each vessel centerline into 50-100 points, calculating the minimal distance to the respective surface ($\Gamma_\mathrm{cap}$ or $\Gamma_\mathrm{vil}$), and taking the mean value of said minimal distances, collapsing every vessel to a single distance value. The distributions of vessel-averaged minimal distances from centerline to capillary surface (red) and to villous surface (blue) are shown for Specimens 1--4; each network comprises between 18 and 43 vessels. The mean and standard deviation for capillary and villous distances of each specimen are given in the figures in the form $(\text{mean} \pm \text{SD})\units{\mu{m}}$. Across all specimens, the vessel-averaged minimal distances from centerline to capillary surface is 8.0$\units{\mu{m}}$, and from centerline to villous surface it is 17.9$\units{\mu{m}}$.}}
\label{fig:histogram}
\end{figure}

The images used here comprise four sets of smooth 3D meshes of fetal vasculature and the accompanying villous membrane (Fig.~\ref{fig:histogram}A), segmented from stained confocal microscopy data (Fig.~1B) as described previously \cite{mayo2016three}. Image dimensions are approximately $(250\times250\times150)\units{\mu{m}}$. 

The watershedding algorithm \textit{AutoSkeleton} of {FEI Amira}$^\text{\texttrademark}$ 6.4  was used to skeletonize capillary centerlines from 3D meshes, as illustrated in Fig.~1D.   Having identified branching points, each network can be represented as a graph (for example, a 2D projection of the 37-segment graph for Specimen 1 is illustrated in the inset to Fig.~\ref{fig:discrete-model}D below).

We extracted geometrical statistics for each capillary branch (capillary length, \chg{and vessel-averaged minimal distances from centerline to capillary surface  and from centerline to villous surface}) using {Wolfram Mathematica}$^\text{\textregistered}$ 11.2.  As Fig.~\ref{fig:histogram}B illustrates,  the \chg{vessel-averaged minimal distances across all specimens from centerline to capillary surface is 8.0$\units{\mu{m}}$, and from centerline to villous surface it is 17.9$\units{\mu{m}}$}.

\section{\label{sec:computational-model}Computational model}

\subsection*{Governing equations}

In simulations, we model fetal blood flow using the Stokes equations 
\begin{equation}
\eta\nabla^{2}\boldsymbol{u}=\nabla p,\qquad\nabla\cdot\boldsymbol{u}=0.
\label{eq:basic-Stokes-PDE}
\end{equation}
Here {$\boldsymbol{u}$ is the fluid velocity field, $p$ the fluid pressure and $\eta$ the dynamic viscosity of fetal blood, which is treated as Newtonian in 3D simulations; we take $\eta = 2\times 10^{-3} \units{Pa\!\cdot\!s}$ (appropriate for blood with 48\% hematocrit in a 20$\units{\mu{m}}$ vessel; see \cite{pearce2016image})}. We address the effects of nonlinear blood rheology in Sec.~\chgrev{\ref{sec:discrete-network-model}} 
below.


The solute concentration $c$ in blood is assumed to obey the linear advection-diffusion equation
\begin{equation}
B\mathbf{u} \cdot \nabla c = D_\text{p} \nabla^2 c, 
\label{eq:basic-advdiff}
\end{equation}
where $D_\text{p}$ is the solute diffusion coefficient in plasma.  The parameter $B=1$ for most solutes, but for species that bind to hemoglobin $B$ quantifies the facilitated transport by red blood cells \cite{serov2015analytical,pearce2016image}.  For example, for oxygen \cite{serov2015analytical,pearce2016image}
\begin{equation}
B = 1 + c_\text{max} K k_\text{hn} \left/ \rho_\text{bl} \right. \approx 141\,,
\label{eq:enhancement}
\end{equation}
where $c_\text{max}$ is the oxygen content of fetal blood at full saturation, $K$ is the gradient of the linearized fetal oxygen-hemoglobin dissociation curve \cite{pearce2016image}, 
$k_\text{hn}$ is the Henry's law coefficient and $\rho_\text{bl}$ is the density of blood. 
In villous tissue, solute transport is governed by the diffusion equation
\begin{equation}
D_\text{t} \nabla^2 c = 0
    \label{eq:laplace}
\end{equation}
where $D_t$ is the solute diffusion coefficient in tissue.   Linearity of Eqs \eqref{eq:basic-advdiff}, \eqref{eq:laplace} is convenient in allowing solute fields to be rescaled to describe transport of solutes with different concentrations.

\subsection*{Boundary conditions}

\begin{figure}[t!]
\centering
\includegraphics[width=1\textwidth]{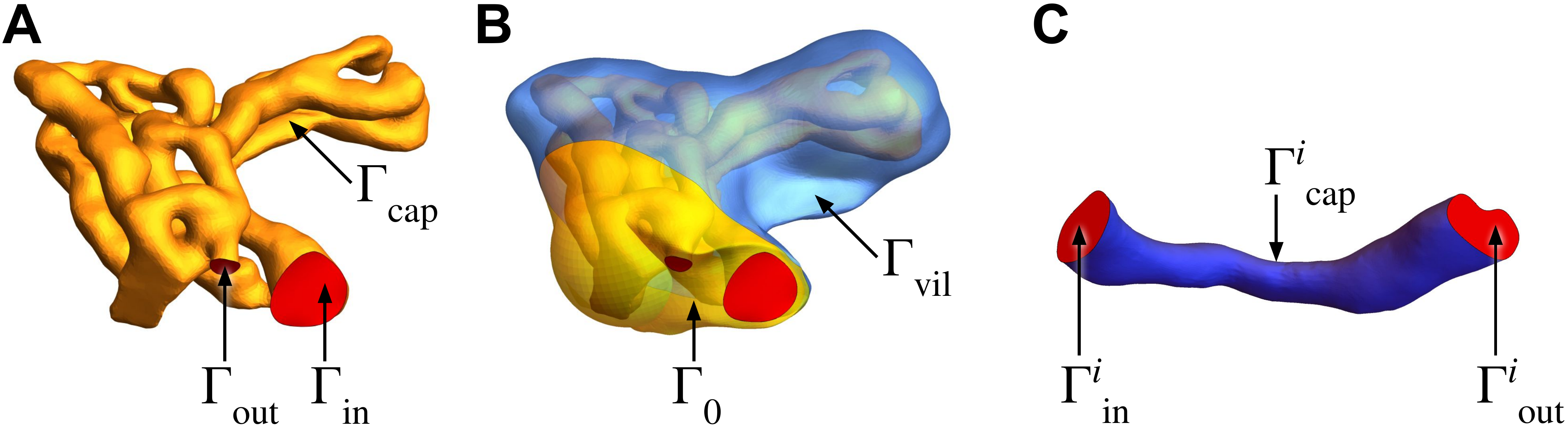}
\caption{Surfaces on which boundary conditions are imposed. Relevant surfaces for the Stokes problem (\textbf{A}, see Eqs \eqref{eq:inletPressure}--\eqref{eq:noSlip}) and additional surfaces for the advection-diffusion solute transport problem and computation of carrying capacity  $N_\text{max}$ (\textbf{B}, see Eqs  \eqref{eq:inletConcentration}--\eqref{eq:carrying-capacity-last}) are shown. When integrating solute transport fluxes over individual capillaries, the labelling convention shown in \textbf{C} is used.}
\label{fig:boundary-conditions}
\end{figure}

The surfaces bounding the domains in which Eqs \eqref{eq:basic-Stokes-PDE}, \eqref{eq:basic-advdiff}, \eqref{eq:laplace} are solved are illustrated in Fig.~\ref{fig:boundary-conditions}A,B.   For the Stokes problem Eq. \eqref{eq:basic-Stokes-PDE}, blood enters through the inlet surface $\Gamma_{\text{in}}$ and leaves via $\Gamma_{\text{out}}$, driven by a pressure difference $\Delta P$ imposed between inlet and outlet. A no-slip condition is imposed on the capillary surface $\Gamma_{\text{cap}}$.   The boundary conditions on the flow are therefore
\begin{align}
p & =\Delta P\quad\text{on}\quad\Gamma_\text{in},\label{eq:inletPressure}\\
p & =0\quad\text{on}\quad\Gamma_\text{out},\label{eq:outletPressure}\\
\boldsymbol{u} & =0\quad\text{on}\quad\Gamma_\text{cap}.\label{eq:noSlip}
\end{align}
Fetal blood is assumed to enter solute-free at the inlet $\Gamma_\text{in}$ and zero diffusive solute flux is imposed at the outlet $\Gamma_\text{out}$.   
Although it is difficult to reliably identify inlet and outlet vessels from the reconstructed geometry alone, the choice has {no} impact on flow resistance {when there is a single inlet and outlet, nor on} the maximum diffusive flux (see below) in the diffusion-limited regime.
The solute concentration and diffusive solute flux are assumed continuous across the internal boundary $\Gamma_{\text{cap}}$.  The maternal solute concentration $c=c_\text{mat}$ is imposed on the villous surface $\Gamma_\text{vil}$ and no diffusive flux is imposed between the inlet/outlet and the villous surface (on $\Gamma_0$) to avoid artificial sharp gradients.  Together, the external boundary conditions on the solute are
\begin{align}
c=0 & \quad\text{on}\quad\Gamma_\text{in},\label{eq:inletConcentration}\\
\boldsymbol{n}\cdot\nabla c=0 & \quad\text{on}\quad\Gamma_\text{out},\,\Gamma_0,\label{eq:outletConcentration}\\
c=c_{\text{mat}} & \quad\text{on}\quad\Gamma_\text{vil}.\label{eq:maternalConcentration}
\end{align}
For oxygen, we assume $c_\text{mat} \approx 0.07 \units{mol/m^3}$ \cite{pearce2016image}.

\subsection*{Net solute transfer}
The net solute transfer rate $N$ of the network is defined as the diffusive flux across $\Gamma_{\text{vil}}$ or equivalently across $\Gamma_{\text{cap}}$.  
As diffusive fluxes across $\Gamma_{\text{in}}$ are very small for the parameters of interest, $N$ is well approximated as the advective flux {leaving the {flow domain} capillary network}
\begin{equation}
    N=\iint_{\Gamma_\text{out}} B\,c\,\boldsymbol{n}\cdot \boldsymbol{u}\, \mathrm{d}A,
    \label{net-uptake-network}
\end{equation}
where $\boldsymbol{n}$ is the unit outward normal to $\Gamma_{\text{out}}$.  We test mass conservation by comparing the advective flux Eq. \eqref{net-uptake-network} over the capillary domain {($\Gamma_\text{out}$)} with the diffusive flux over the villous domain {($\Gamma_\text{in}$ and $\Gamma_\text{vil}$)} to validate the numerical implementation.

\subsection*{The maximum diffusive flux}
The maximum diffusive flux (or carrying capacity) $N_\text{max}$ corresponds to the net solute flux arising when the flow is sufficiently strong for the inlet condition $c=0$ to apply across $\Gamma_\text{cap}$.   It can be calculated by solving Eq. \eqref{eq:laplace} over the villous tissue domain with boundary conditions 
\begin{align}
c=0 & \quad\text{on }\quad
\Gamma_{\text{cap}}\label{eq:carrying-capacity-first}\\
\boldsymbol{n}\cdot\nabla c=0 & \quad\text{on }\quad\Gamma_{0},\\
c=c_{\text{mat}} & \quad\text{on }\quad\Gamma_{\text{vil}}.\label{eq:carrying-capacity-last}
\end{align}
and evaluating {the diffusive flux across the capillary surface}
\begin{equation}
    N_\text{max}=-\iint_{\Gamma_\text{cap}} D_\text{t}\, \boldsymbol{n}\cdot \nabla c\,\, \mathrm{d}A\,.
\end{equation}
The parameter characterizing integrated exchange area over exchange distance is then defined by 
\begin{equation}
\mathcal{L}=N_{\text{max}}\,/\left( D_\text{t}\,c_{\text{mat}}\right).
\label{eq:ell}
\end{equation}

\subsection*{Numerical implementation}
We used {COMSOL Multiphysics}$^\text{\textregistered}$ 5.3a to solve the coupled flow and transport problems defined above.  
For the Stokes problem {in Figs 4, \chgrev{5A,B} and S3}, we used the \textit{Creeping Flow} module, calculating the solution on capillary meshes of approximately {5.6} million tetrahedral elements. To calculate the concentration field {in these figures}, we used the \textit{Transport of Diluted Species} module on meshes of approximately {61.6} million tetrahedral elements.  \chgrev{To ensure} that concentration boundary layers (should they arise) and fine details of the mesh (such as local near-contact of villous and capillary meshes) were resolved, \chgrev{we performed a mesh convergence analysis. For the most intricate mesh (Specimen 3), an almost nine-fold increase from approximately 7.6 to 65.6 million tetrahedral elements changed the net uptake flux $N$ at $\Delta P = 40 \units{Pa}$ 
by less than 3\%}.

\begin{figure}[t!]
\centering
\includegraphics[width=1\textwidth]{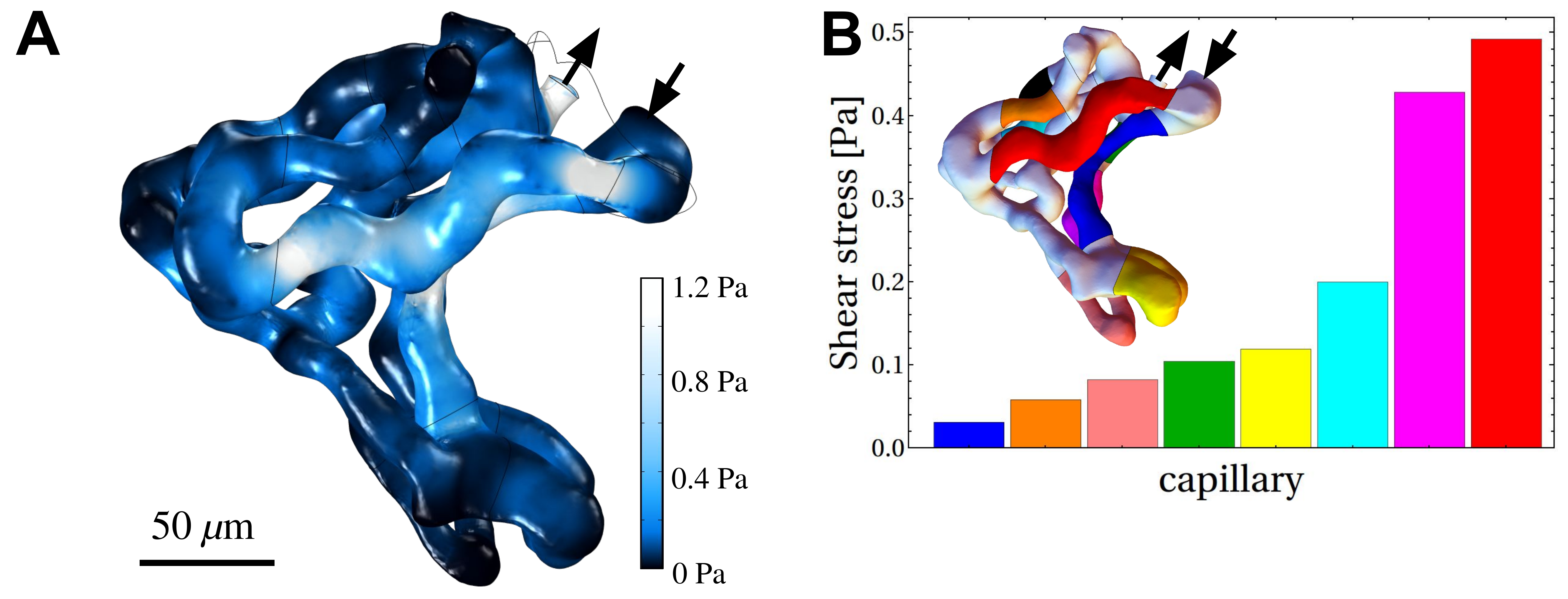}
\caption{\textbf{A} Predicted shear stresses in Specimen 1 shown for an inlet-outlet pressure drop of $\Delta P = 40\units{Pa}$. The highest wall shear stresses (white, around $1.2\units{Pa}$) occur where capillaries are thin and flow speeds are greatest, e.g. at the outlet (near the outward-pointing arrow). \textbf{B} Average shear stresses of the nine capillaries (inset) discussed in Fig. 4A--C in the main text.}
\label{fig:shear-stress}
\end{figure}

To calculate fluxes, we used \emph{Accurate Fluxes} in \text{COMSOL} (\text{tds.ncflux\_c} and \text{tds.ndflux\_c} for advective and diffusive fluxes, respectively) and ensured that advective fluxes integrated over closed domains match.  In doing so, we evaluated the net flux over an individual capillary $i$ (see Fig.~\ref{fig:boundary-conditions}C) using
\begin{equation}
    N_i=\iint_{\Gamma_\text{out}^i} B\,c\,\boldsymbol{n}\cdot \boldsymbol{u}\, \mathrm{d}A - \iint_{\Gamma_\text{in}^i} B\,c\,\boldsymbol{n}\cdot \boldsymbol{u}\, \mathrm{d}A,
\end{equation}
and evaluated $N^i_\text{max}$ for an individual capillary via the diffusive flux over its capillary wall $\Gamma^i_\text{cap}$. {In the case of net flux computations over the entire network (Fig. 2 in the main text), we calculate $N$ by integrating over the entire villous surface to minimise error introduced by very small outlet surfaces. }

{The data underlying Figs 2 and 3 of the main text {were} produced with {comparable} mesh quality for all four {specimens}. Stokes flow on the capillary domain was solved on {meshes with between 0.3 to 1.2 million tetrahedral elements}. Transport was solved on the villous and capillary domains on {meshes with between 4.1 and 22.4 million tetrahedral elements. For Specimen 1, the comparison between the net solute transfer across the network calculated at high resolution (61.6 million tetrahedral element mesh for transport problem, as used in Fig. 4) and low resolution (15.0 million tetrahedral element mesh for transport problem, as used in Fig. 2) led to a maximal relative error of 5.2\% at a very high pressure drop ($\Delta P = 2725.2 \units{Pa}$); at a physiological pressure drop of $\Delta P = 40 \units{Pa}$, the relative error was 0.6\%.} }
%


\subsection*{Inlets, outlets and boundary surfaces}
The three-dimensional mesh data of capillary and villous surfaces has a number of imperfections {and imaging artefacts} that add a subjective component to the identification of boundary conditions. The {S}pecimen 1-4 meshes have between three and five candidate locations for inlets and outlets, and we made our choice of inlet and outlet on a case-by-case basis: {i}n {S}pecimen 1, 3 and 4 we identified one likely inlet and one likely outlet per specimen, and made slight modifications {by locally adding small hemispheres} to the villous surface at the discarded inlet/outlet candidate locations. These modifications ensure that apart from at the inlet and outlet, the villous surface does not come {unnecessarily} close to the capillary surface.
Another imperfection of the {imaging} data arose due to the depth limitations of confocal microscopy, which 
sometimes makes it unclear if a part of the villous surface was originally in contact with maternal blood or resulted from an artificial{ly} cut-off {internal boundary} on which {unphysiological} oxygen exchange could occur. {In the latter} case, a no-flux boundary condition {is applied as} appropriate ($\Gamma_0$ in Fig. \ref{fig:boundary-conditions}). We identified the no-flux planes vs. exchange planes according to our best judgement. Comparisons between different choices of no-flux planes revealed differences in $\mathcal{L}$ of up to 12\%. 

\subsection*{Shear stress distributions}
In addition to the results reported in Figs~2--4 of the main text, the computational model provides detailed maps of predicted shear stress within capillaries (Fig.~\ref{fig:shear-stress}A).  For a network pressure drop $\Delta P$ of $40\units{Pa}$, the shear stress is everywhere below a maximum of approximately $1.2\units{Pa}$; for comparison, Olesen \hbox{et al.} \cite{Olesen88} estimated a physiological shear stress range between $0.5\units{Pa}$ to $2\units{Pa}$ in arterioles of comparable diameter to those encountered here.  The shear stress at any location within the network is linearly proportional to $\Delta P$ under a Newtonian Stokes flow approximation, suggesting that an increase of $\Delta P$ to around $100\units{Pa}$ remains within a physiological range.  Regions of locally elevated shear stress are found at constrictions and in vessels carrying greater flow, for example near the inlet or outlet.  The variation in average shear stress between vessels (Fig.~\ref{fig:shear-stress}B) is notable\chgrev{, indicating local variations in flow resistance. However, these results depend on the specimen fixation pressure and the choice of flow rheology model (see Sec.~\ref{sec:discrete-network-model} below and Figs~\ref{fig:individual-capillaries} and \ref{fig:discrete-model} in the main text.)}.

\subsection*{Model parameters for passively transported solutes}
The developed framework readily extends to a variety of relatively small and mobile solutes. Table~\ref{tab:different-solutes} summarises and estimates key transport parameters{, specifically} effective advection-enhancement factors $B$, plasma $D_\text{p}$ and tissue $D_\text{t}$ diffusivities, as well Damk\"ohler and diffusive capacity numbers ($\mathrm{Da}_\text{rel}$ and $\mu_\text{rel}$) relative to oxygen values.

\section{\label{sec:single-capillary-regression}Transport in a single cylindrical capillary}

We now motivate the form of the regression equation, Eq. (3) in the main text, by analysing transport in a single capillary.  We assume axisymmetry, denoting parameters in this special case with a circle superscript. 

\newcommand{\twocolumncentral}[1]{\multicolumn{2}{c}{$\quad #1\hfill$}}
\setlength{\heavyrulewidth}{.08em}
\setlength{\lightrulewidth}{.05em}
\setlength{\defaultaddspace}{.3em}
\begin{table*}[t!]
\centering
\begin{tabular}{c c c c c c}
\toprule
\multirow{2}{*}{Solute} & \multirow{2}{*}{$B$} & $D_\text{p}$ & $D_\text{t}$ & 
\multirow{2}{*}{$\left(\mathrm{Da}_\text{rel}\right)^{-1}$} & \multirow{2}{*}{$\mu_\text{rel}$} \tabularnewline
           & & \multicolumn{2}{c}{$\times\,10^{-9}\units{\left[m^2/s\right]}$} & & \\
\midrule  \addlinespace
carbon monoxide (CO) & $\sim10^4\,^a$ & \twocolumncentral{2\,^b} & $\sim 10^2$ & 1 \tabularnewline
mannitol             & 1 & \:0.7$\,^c$ & $\sim\!\left(10^{-4}-10^{-3}\right)^d$ & $\sim10-10^{2}$ & $\sim10^{-3}-10^{-2}$ \tabularnewline
fructose             & 1 & \:0.7$\,^e$ & $\sim\!\left(10^{-4}-10^{-3}\right)^f$ & $\sim10-10^{2}$ & $\sim10^{-3}-10^{-2}$ \tabularnewline
glucose              & 1 & \:0.7$\,^e$ & $\sim\!\left(10^{-3}-10^{-2}\right)^f$ & $\sim 1 - 10$   & $\sim10^{-2}-10^{-1}$ \tabularnewline\addlinespace
oxygen (O$_2$)       & $\approx 140\,^g$ & \twocolumncentral{2\,^b} & $1$             & 1 \tabularnewline\addlinespace
carbon dioxide (CO$_2$)  & $\sim\!\left(1-10\right)^{\,h}$ & \twocolumncentral{1.9\,^b}  & $\sim10^{-2}-10^{-1}$   & 1 \tabularnewline\addlinespace
nitrous oxide (N$_2$O)  & 1 & \twocolumncentral{2.6\,^c}  & $\sim 10^{-2}$  & 1 \tabularnewline\addlinespace
urea                    & 1 & \twocolumncentral{1.4\,^b}  & $\sim 10^{-2}$  & 1 \tabularnewline\addlinespace
ethanol                 & 1 & \twocolumncentral{1.2\,^e}  & $\sim 10^{-2}$  & 1 \tabularnewline\addlinespace
caffeine                & 1 & \twocolumncentral{0.8\,^c}  & $\sim 10^{-2}$  & 1 \tabularnewline
\bottomrule
\end{tabular}
\caption{
Characteristic parameters for various passively transported solutes. The constant $B$ describes the solute carrying capacity by the red blood cells. 
The solute diffusivities in blood plasma and in villous tissue (where the solute is dissolved in water) are $D_\mathrm{p}$ and $D_\mathrm{t}$ respectively. The Damk\"ohler and diffusive capacity numbers relative to oxygen values are $\mathrm{Da}_{\text{rel}} \equiv D_\mathrm{t}^\text{solute} B^\text{oxygen}/(D_\mathrm{t}^\text{oxygen} B^\text{solute})$ and $\mu_{\text{rel}} \equiv D_\mathrm{t}^\text{solute} / D_\mathrm{p}^\text{solute}$ respectively.
Data are taken (with most $D_\mathrm{p}$ values given in literature at 25\textcelsius) or estimated from
$^a$\cite{Faber95,Longo77},          
$^b$\cite{Cussler_Book09},           
$^c$\cite{Hills2011}.                
$^d$\cite{Bain88},                   
$^e$\cite{CRC_Handbook14,Ribeiro06}, 
$^f$\cite{Barta10,Holmberg56,LevkovitzElad13},
$^g$\cite{pearce2016image},          
$^h$\cite{Longo73}.                  
}
\label{tab:different-solutes}
\vspace{-1em}
\end{table*}



\begin{figure}[t]
\centering
\includegraphics[width=0.45\textwidth]{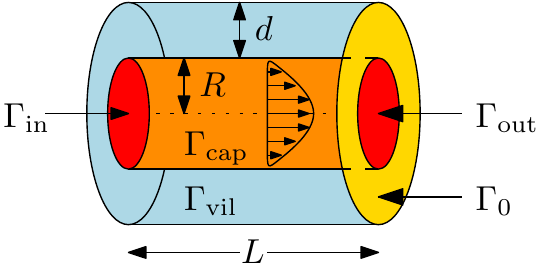}
\caption{A capillary is represented in the discrete network model by a cylinder with Poiseuille flow and a surrounding cylindrical shell representing the villous domain. The boundary surfaces are labeled in the convention of Fig.~\ref{fig:boundary-conditions}.}
\label{fig:virtualcapillary}
\end{figure}

Consider a cylindrical feto-placental capillary of length $L$
and radius $R$ within an annular villous volume of thickness $d$ (Fig.~\ref{fig:virtualcapillary}).  In cylindrical coordinates, the flow problem Eq. \eqref{eq:basic-Stokes-PDE}, \eqref{eq:inletPressure}--\eqref{eq:noSlip} 
has the familiar Poiseuille solution for the axial velocity
\begin{equation}
u\left(r\right)=u_{\text{max}}\left(1-\frac{r^{2}}{R^{2}}\right),\quad u_{\text{max}}=\frac{\Delta PR^{2}}{4\eta L}.
\end{equation}
The cross-sectionally averaged velocity is $\langle u\rangle=\frac{1}{A}\iint u\,\text{d}A=u_\text{max}/2$, where $A$ is the cross-section area. The volume flux $q=\iint u\,\text{d}A$ is related to the pressure drop $\Delta P$ across the capillary via $\Delta P = \mathcal{R}^\circ q$ with the Poiseuille resistance 
\begin{equation}
    \mathcal{R}^{\circ}=\frac{8 \eta L}{\pi R^4}.
    \label{eq:resistance-tube}
\end{equation}
The advection-diffusion problem {given by Eqs} \eqref{eq:basic-advdiff}, \eqref{eq:inletConcentration}--\eqref{eq:maternalConcentration} becomes 
\begin{eqnarray}
Bu\frac{\partial c}{\partial z} & = & D_\text{p}\left[\frac{1}{r}\frac{\partial}{\partial r}\left(r\frac{\partial c}{\partial r}\right)+\frac{\partial^{2}c}{\partial z^{2}}\right] \label{eq:advection-diffusion-1}\\
0 & = & D_\text{t}\left[\frac{1}{r}\frac{\partial}{\partial r}\left(r\frac{\partial c_\text{t}}{\partial r}\right)+\frac{\partial^{2}c_\text{t}}{\partial z^{2}}\right]\label{eq:advection-diffusion-2}.
\end{eqnarray}
Here we use $c_\text{t}$ to denote the solute concentration in villous tissue. The boundary conditions are:
\begin{alignat}{2}
c & =0 &\quad \text{at }z & =0\\
\frac{\partial c}{\partial z} & =0 & \text{at }z & =L\\
c_\text{t} & =c_\text{mat} & \text{at }r & =R+d\\
D_\text{p}\frac{\partial c}{\partial r} & =D_\text{t}\frac{\partial c_\text{t}}{\partial r}\label{eq:Neumann} & \text{at }r & =R\\
c & =c_\text{t} & \text{at }r & =R\\
\frac{\partial c}{\partial r} & =0 & \text{at }r & =0.
\end{alignat}
Neglecting axial diffusion in Eq. \eqref{eq:advection-diffusion-2}, we can obtain an explicit solution for $c_\text{t}$ in terms of $c$, allowing us to write the full problem in terms of the capillary concentration $c$ only. The Neumann condition Eq. \eqref{eq:Neumann} becomes a Robin condition 
\begin{equation}
\frac{\partial c}{\partial r}=\mu_\circ\left(\frac{c_\text{mat}-c}{R}\right)\quad\text{at }r=R
\label{eq:single-capillary-flux}
\end{equation}
with the diffusive capacity
\begin{equation}
    \mu_\circ = \frac{D_\mathrm{t}/D_\mathrm{p}}{\log(1+d/R)}.
    \label{eq:mu-single-capillary}
\end{equation}
Setting $c=0$ in Eq.~\chgrev{\eqref{eq:single-capillary-flux}} 
and integrating the diffusive flux over $\Gamma_{\text{cap}}$, it follows that $N_{\text{max}}^\circ=2\pi D_\text{t}\, c_{\text{mat}}\,L/\log(1+d/R)$.

\subsection*{Asymptotic approximation}
Introducing the non-dimensional variables
\begin{equation}
\Pi=\frac{B\,R\,u_{\text{max}}}{\sqrt{\mu_\circ}D_\mathrm{p}},\quad\alpha=\frac{L}{R},\quad\hat{r}=\frac{r}{R},\quad\hat{z}=\frac{z}{R},\quad\hat{c}=\frac{c}{c_\text{mat}},
\label{eq:non-dimensional-variables}
\end{equation}
the problem is specified in terms of $\mu_\circ$, a modified P\'eclet number $\Pi$ and the tube aspect ratio $\alpha$ as
\begin{equation}
\Pi\sqrt{\mu_0}\left(1-\hat{r}^{2}\right)\frac{\partial\hat{c}}{\partial\hat{z}}=\frac{1}{\hat{r}}\frac{\partial}{\partial\hat{r}}\left(\hat{r}\frac{\partial\hat{c}}{\partial\hat{r}}\right)+\frac{\partial^{2}\hat{c}}{\partial\hat{z}^{2}}
\label{eq:transport-nondimensional}
\end{equation}
with boundary conditions 
\begin{alignat}{2}
\hat{c} & =0 &\quad \text{at }\hat{z} & =0\label{eq:bc-nondim-1}\\
\frac{\partial\hat{c}}{\partial\hat{r}} & =0 & \text{at }\hat{r} & =0\\
\frac{\partial\hat{c}}{\partial\hat{z}} & =0 & \text{at }\hat{z} & =\alpha\\
\frac{\partial\hat{c}}{\partial\hat{r}} & =\mu_\circ (1-\hat{c}) & \text{at }\hat{r} & =1\label{eq:bc-nondim-4}.
\end{alignat}
We now demonstrate how diffusion-limited and strongly or weakly flow-limited regimes can be obtained from this boundary-value problem. 

When the diffusive capacity is low ($\mu_\circ \ll 1$), radial diffusion over a long domain suppresses transverse concentration gradients. 
Following \cite{Woollard2009}, we scale the axial coordinate by $\sqrt{\mu_\circ}$ and approximate the concentration profile as well-mixed, using
\begin{equation}
\hat{z}=\frac{\overline{z}}{\sqrt{\mu_\circ}},\qquad\hat{c}=\hat{c}_{0}\left(\overline{z}\right)+\mu_\circ\hat{c}_{1}\left(\overline{z},\hat{r}\right)+\mathcal{O}\left(\mu_\circ^{2}\right).
\label{eq:scaling-flat-profile}
\end{equation}
The non-dimensional problem Eq. \eqref{eq:transport-nondimensional} then becomes 
\begin{equation}
\Pi\left(1-\hat{r}^{2}\right)\frac{\partial\hat{c}_{0}}{\partial\overline{z}}=\frac{1}{\hat{r}}\frac{\partial}{\partial\hat{r}}\left(\hat{r}\frac{\partial\hat{c}_{1}}{\partial\hat{r}}\right)+\frac{\partial^{2}\hat{c}_{0}}{\partial\overline{z}^{2}}.
\label{eq:transport-flat-profile}
\end{equation}
Integrating Eq. \eqref{eq:transport-flat-profile} over the cross-section and imposing boundary conditions we obtain the ordinary differential equation
\begin{equation}
\frac{\Pi}{4}\hat{c}_{0}'\left(\overline{z}\right)=\tfrac{1}{2}\hat{c}_{0}''\left(\overline{z}\right)+1-\hat{c}_{0}\left(\overline{z}\right),\quad\hat{c}_{0}\left(0\right)=0,\quad c_{0}'\left(\overline{\alpha}\right)=0,
\end{equation}
where $\overline{\alpha}=\sqrt{\mu_\circ}\alpha$. The solution $\hat{c}_0$ to this boundary value problem can be integrated as $N \propto \mu_\circ \left(\int_{\hat{z}=0}^{\alpha}\hat{c}_{0}\mathrm{d}\hat{z}- \alpha\right)$ to find the net uptake. When axial diffusion is weak ($\Pi \gg 1$), we find
\begin{equation}
N\approx N_{\text{FL-DL}}\equiv N_\text{max}^\circ\, \mathrm{Da}_\circ^{-1} \left(1-e^{-\mathrm{Da}_\circ}\right)
\label{eq:nfldl}
\end{equation}
where the relevant inverse Damköhler number is \begin{equation}
    \mathrm{Da}_\circ^{-1}=\frac{D_\mathrm{t}}{D_\mathrm{p}}\frac{\Pi}{4\sqrt{\mu_\circ} \alpha}.
\end{equation} 
Eq.~\eqref{eq:nfldl} encompasses the strongly flow-limited regime $N\approx N_\text{max}^\circ \text{Da}_\circ^{-1}$ when $\text{Da}_\circ\gg 1$ and the diffusion-limited regime $N\approx N_\text{max}^\circ$ when $\text{Da}_\circ\ll 1$.  

In the strongly flow-limited regime, the assumption of a nearly flat concentration profile Eq. \eqref{eq:scaling-flat-profile} is no longer viable, as concentration boundary layers form in a corner region near the inlet of the tube. Instead the L\'ev\^eque approximation must be employed, which requires a transformation into the boundary layer coordinate system \cite{Cussler_Book09,Woollard2009}. This allows us to recover the weakly flow-limited regime
\begin{equation}
    N_{\text{WFL}}=N_{\text{max}}^\circ \alpha_\text{c}\text{Da}_\circ^{-1/3}\mu_\circ^{-2/3}
\end{equation}
where $\alpha_\text{c}\approx 5.5$.  An approximation for $N$ across all physical regimes can then be obtained from a harmonic mean of $N_\text{FL-DL}$ and $N_\text{WFL}$
\begin{equation}
    N^{-1} = N_\text{FL-DL}^{-1}+N_\text{WFL}^{-1}.
    \label{eq:regression-single-tube}
\end{equation}
This predicts $N$ in terms of the geometric parameters $R$, $d$ and $L$, the material parameters $\eta$, $B$, $D_\text{p}$ and $D_\text{t}$, the imposed pressure drop $\Delta P$ and the concentration difference $c_{\text{mat}}$.  The empirical regression equation Eq. (3) in the main text, generalizes this approach to the whole network.

\section{\label{sec:discrete-network-model}A discrete model for transport in a capillary network}

In order to explore the effect of hematocrit on solute transport in feto-placental capillary networks, and to test the system for sensitivity to occlusion of single vessels, we develop a discrete network model that resolves individual capillaries as elements of a graph.  We approximate solute transport by adapting the modified Krogh cylinder formulation in Sec.~\ref{sec:single-capillary-regression} above, ensuring conservation of fluid and solute at all nodes in the capillary network.  We test the reduction from a continuous formulation using partial differential equations (Sec.~\ref{sec:computational-model}) to a discrete (algebraic) representation before using the simplified model to evaluate the distribution of hematocrit in the network, calculated using the empirical law for plasma skimming from \cite{pries1990blood}. The distribution of hematocrit is used to calculate the effective viscosity in each vessel due to the F{\aa}hr{\ae}us--Lindqvist effect.   We also test the sensitivity of the network to blockage of individual vessels.

\subsection*{The capillary network as a directed graph}
Our low-order model for transport in a capillary network adapts and expands Strang's treatment of electrical circuits \cite{Strang1988}.  Consider a network having $m$ segments (capillaries), each with an assigned orientation, and $n$ nodes. To describe the relationship between nodal and segmental quantities, we introduce the $m \times n$ incidence matrix $\mathsf{A}$. Its entries ${A}_{ij}$ are either $+1$, $0$ or $-1$, where $0$ means that an edge and a node are not incident, $+1$ means that a directed edge points towards the node, $-1$ means that the edge points away from the node.  It is helpful to introduce the downstream incidence matrix $\mathsf{A}_+$ (in which all negative entries of $\mathsf{A}$ have been set to zero) and the upstream incidence matrix $\mathsf{A}_-$ (in which all positive entries of $\mathsf{A}$ have been set to zero) such that $\mathsf{A}=\mathsf{A}_- +\mathsf{A}_+$.  Over all the segments we define a vector of scalar fluxes $\boldsymbol{q}=(q_1,\ldots,q_m)^T$, where $q_i>0$ indicates that the flow direction in segment $i$ matches the orientation of the segment $i$.   Over the nodes we define vectors of scalar pressures $\boldsymbol{p}=(p_1,\ldots,p_n)^T$ and inlet concentrations $\boldsymbol{c}=(c_1,\ldots,c_n)^T$.   $\mathsf{A}\boldsymbol{p}$ is then a vector of pressure differences, defined over directed segments.

Writing $\mathsf{B}=\text{diag }(B_1,\ldots,B_m)$ as a diagonal matrix of advection boost coefficients and $\mathsf{Q}=\text{diag }(q_1,\dots,q_m)$ as a diagonal matrix of fluxes, we introduce the $m$-dimensional vector of advective fluxes over segments
\begin{equation}
    \boldsymbol{n}^a = - \mathsf{B}\,\mathsf{Q}\,\mathsf{A}_{-}\boldsymbol{c}.
\end{equation}
Defining transmural fluxes $N_i$ for $i=1,\ldots,m$ using the single tube results Eq. \eqref{eq:regression-single-tube} (in terms of three geometric parameters for each vessel and the pressure drop across it), we construct the diagonal matrix $\mathsf{N} = \text{diag }(N_1,\ldots,N_m)$.   Rescaling the fluxes to the relevant local concentrations, the $m$-dimensional vector of transmural diffusive fluxes is then 
\begin{equation}
    \boldsymbol{n}^{d}=\mathsf{N}\left(\boldsymbol{1}+c_{\text{mat}}^{-1}\mathsf{A}_{-}\boldsymbol{c}\right)
    \label{eq:diffusive-flux-complete-mixing}
\end{equation}
where $\boldsymbol{1}=(1, \ldots, 1)^T$ is an $m$-dimensional vector. 

The discrete flow and transport problem over the network can then be written compactly as
\begin{alignat}{2}
\text{volume flux conservation} &  & \mathsf{A}^{T}\boldsymbol{q} & =\boldsymbol{f}_{\text{ext}}\label{eq:discrete-volume-flux}\\
\text{flow resistance} &  & \mathsf{A}\boldsymbol{p}-\mathsf{R}\boldsymbol{q} & =\boldsymbol{0}\label{eq:discrete-Poiseuille}\\
\text{advection-diffusion transport} &  &\quad \mathsf{A}^{T}\boldsymbol{n}^{a}+\mathsf{A}_{+}^{T}\boldsymbol{n}^{d} & =\boldsymbol{g}_{\text{ext}}.\label{eq:discrete-Transport}
\end{alignat}
Here $\boldsymbol{f}_\text{ext}=(-Q_\text{ext},0,\ldots,0, Q_\text{ext})$ is an $n$-dimensional vector having first and last entries accounting for the scalar volume flux $Q_\text{ext}$ entering and leaving the system.  The system Eq. \eqref{eq:discrete-volume-flux} of $n$ linear equations enforces conservation of volume flux at every node, which is coupled to the $m$-dimensional linear system Eq. \eqref{eq:discrete-Poiseuille} describing Poiseuille's Law $\Delta P = \mathcal{R}^\circ q$ for the network; here $\mathsf{R}=\text{diag}(\mathcal{R}^\circ_1,\ldots,\mathcal{R}^\circ_m)$.  Finally, the $m$-dimensional linear system Eq. \eqref{eq:discrete-Transport} describes the transport, i.e. the balance of advective and diffusive fluxes. In total there are $n+2m$ linear equations for the unknowns
$\boldsymbol{p}$, $\boldsymbol{q}$ and $\boldsymbol{c}$. Boundary conditions can be imposed through the $n$-dimensional vectors $\boldsymbol{f}_\text{ext}$, $\boldsymbol{g}_\text{ext}$. 
A key assumption of the model is that concentration is fully mixed (i.e. has a radially independent profile) at every node. 

\chgrev{In Fig.~\ref{fig:discrete-model}C, we compare the computational results of the discrete network model with the regression Eq.~\eqref{eq:regression} applied to the whole network, which depends on the maximum achievable uptake flux $N_\mathrm{max}$ and flow resistance $\mathcal{R}$ of the discrete network. These quantities were computed directly from the discrete model (parameterized directly by the vessel-averaged statistics, Fig.~\ref{fig:histogram}, rather than computational fluid dynamics results). We estimate $\mathcal{R}$ by calculating the flow rate at the inlet segment of the discrete network (which is equal to the flow rate at the outlet segment) and dividing by the applied inlet-outlet pressure drop. To compute $N_\mathrm{max}$, we apply a sufficiently high inlet-outlet pressure drop $\Delta P$ such that further increase in $\Delta P$ does not change the net uptake $N$ by more than  0.01\%, which is then used as the discrete network's $N_\mathrm{max}$. }


\subsection*{Hematocrit and nonlinear rheology}

Having established that the discrete \chgrev{network} model provides a reliable representation of transport at the level of individual vessels \chgrev{(Fig.~\ref{fig:discrete-model}A,B)}, we now use it to explore the impact of nonlinear blood \chgrev{rheology} on solute transport.
\chgrev{This model accounts for plasma skimming, the F{\aa}hr{\ae}us effect and the F{\aa}hr{\ae}us--Lindqvist effect; to incorporate facilitated transport, we use a linearized oxygen-hemoglobin dissociation curve \cite{pearce2016image,serov2015analytical}}.

The distribution of hematocrit in the network is calculated using the empirical law for plasma skimming from \cite{pries1990blood}.   The fraction of hematocrit $FQ_E$ entering a vessel at a bifurcation is found in terms of the fraction of blood flow $FQ_B$ entering that vessel using
\begin{equation}
\text{logit}~ FQ_E = C_1+C_2~\text{logit}\left(\frac{FQ_B-X_0}{1-2X_0}\right), \label{eq:plasmaskimming}
\end{equation}
where $\text{logit}~x \equiv \ln \left(x / \left(1-x\right)\right)$ and the parameter $X_0$ defines the minimal
fractional blood flow required to draw red blood cells into the branch. The constants in Eq. \eqref{eq:plasmaskimming} are given by
\begin{align}
C_1 = -6.96 \ln \left(\frac{R_1}{R_2}\right) / \left(2 R_F\right), \quad C_2 = 1+ 6.98 \left(\frac{1-H_F}{2R_F}\right),\quad X_0 = \frac{0.4}{2 R_F},
\end{align}
where $H_F$ and $R_F$ are the hematocrit and the radius of the feeding vessel, $R_1$ is the radius of the vessel being considered and $R_2$ is the radius of the other vessel in the bifurcation (radii are measured in $\units{\mu m}$).

The distribution of hematocrit $H$ is used to calculate the effective viscosity $\eta$ in each vessel due to the F{\aa}hr{\ae}us--Lindqvist effect according to
\begin{equation}
\eta \left/ \eta_\text{p} \right. = 1 + \frac{e^{H \beta} - 1}{e^{0.45 \beta} - 1} \left(110 e^{- 2.848 R} + 3 - 3.45 e ^ {-0.07 R} \right) 
\label{eq:fleffect},
\end{equation}
where $\beta=4 / \left(1+ \exp{\left(-0.0593\left(2 R - 6.74\right)\right)}\right)$ and $\eta_\text{p}=\chgrev{10^{-3}\units{Pa\!\cdot\!s}}$ is the viscosity of plasma 
\chgrev{($\eta\approx2\eta_\text{p}$ in a vessel of radius $R=10\units{ \mu m}$ for $H=0.48$)}
\cite{pries1990blood}. 
The two steps above are implemented in the discrete model and iterated using a custom MathWorks MATLAB$^\text{\textregistered}$ R2016a code until the solution no longer changes, typically after less than 50 iterations. The MATLAB code was coupled with Wolfram Mathematica$^\text{\textregistered}$~11.2 via the MATLink 1.1 package.


\end{document}